\documentclass{aa}  
\usepackage{comment}
\usepackage{afterpage}
\usepackage{pdflscape}
\usepackage{graphicx}
\usepackage{subcaption}
\usepackage{booktabs}
\usepackage{amsmath}
\usepackage{pdfpages}
\usepackage{wrapfig}
\usepackage{txfonts}
\usepackage{float}
\usepackage{caption}
\usepackage{calrsfs}
\DeclareMathAlphabet{\pazocal}{OMS}{zplm}{m}{n}
\usepackage[colorlinks=true, linkcolor=blue, citecolor=blue, urlcolor=blue]{hyperref}
\usepackage{academicons}
\usepackage{xcolor}
\usepackage{lineno}
\usepackage[normalem]{ulem}
\usepackage{natbib}
\bibpunct{(}{)}{;}{a}{}{,} 

\newcommand{\orcid}[1]{\href{https://orcid.org/#1}{\textcolor{orange}{\includegraphics[scale=0.008]{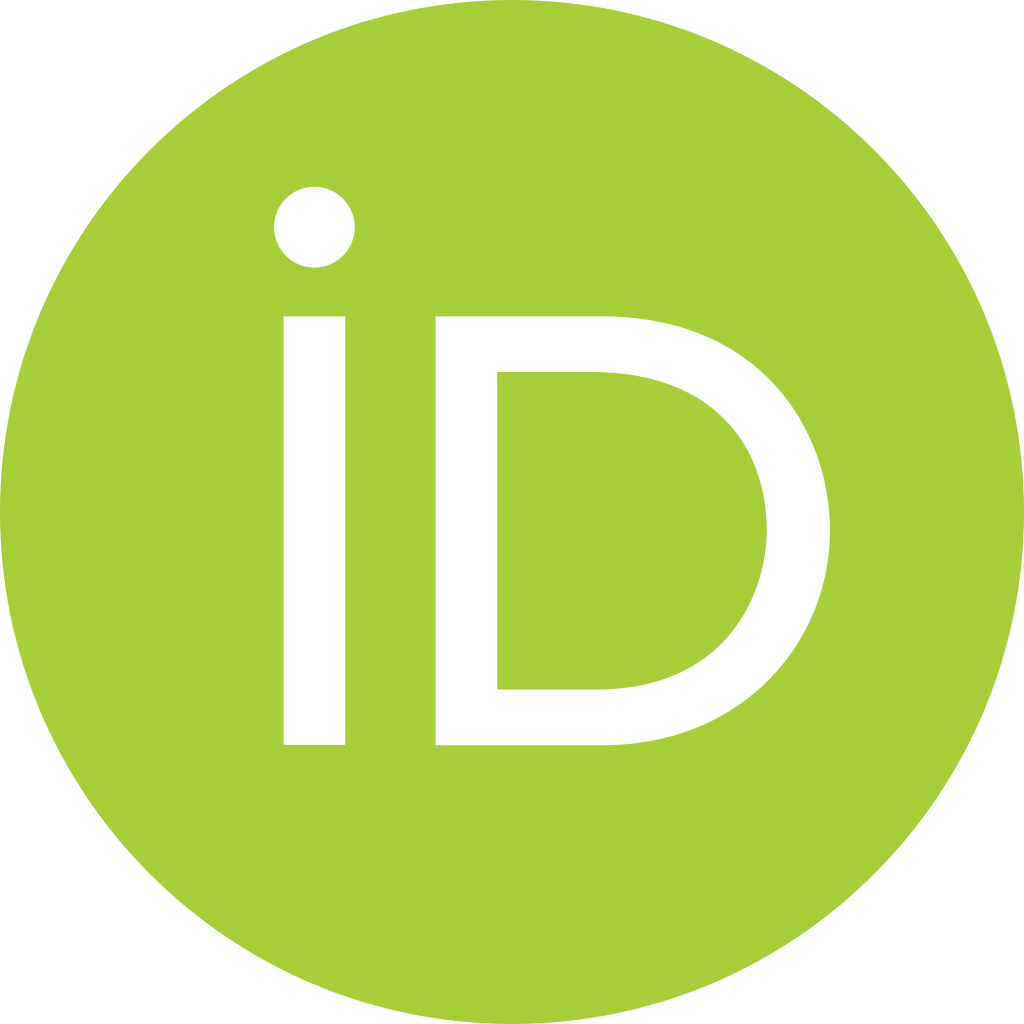}}}}

\begin{document}

   \title{CORALIE radial-velocity search for companions\\ around evolved stars (CASCADES)}
   \subtitle{IV: New planetary systems around HD 87816, HD 94890, and HD 102888 and an update on HD 121056}

   \author{E. Fontanet\inst{1}\orcid{0000-0002-0215-4551},
        S. Udry\inst{1}\orcid{0000-0001-7576-6236},
        D. Ségransan\inst{1}\orcid{0000-0003-2355-8034},
        P. Figueira\inst{1}\orcid{0000-0001-8504-283X},
        J.~A.~Acevedo~Barroso\inst{2}\orcid{0000-0002-9654-1711},
        B. Akinsanmi\inst{1}\orcid{0000-0001-6519-1598},
        M. Attia\inst{1}\orcid{0000-0002-7971-7439},
        M. Battley\inst{1,3}\orcid{0000-0002-1357-9774},
        S. Bhatnagar\inst{1,4}\orcid{0000-0001-9178-168X},
        M. Bugatti\inst{1},
        Y. Carteret\inst{1}\orcid{0000-0002-6159-6528},
        H. Chakraborty\inst{1}\orcid{0000-0002-5177-1898},
        A. Deline\inst{1},
        C. Farret Jentink\inst{1}\orcid{0000-0001-9688-9294},
        Y.G.C. Frensch\inst{1}\orcid{0000-0003-4009-0330},
        M. Houelle\inst{1},
        B. Lavie\inst{1}\orcid{0000-0001-8884-9276},
        C. Lovis\inst{1}\orcid{0000-0001-7120-5837},
        M. Mayor\inst{1},
        A. Nigioni\inst{1}\orcid{0009-0004-5882-6574},
        G. Ottoni\inst{1}\orcid{0000-0001-9305-9631},
        F. Pepe\inst{1},
        D. J. M. Petit dit de la Roche\inst{1}\orcid{0000-0002-8963-3810},
        M. Shinde\inst{1}\orcid{0000-0002-8024-3779},
        N. C. Santos \inst{5,6}\orcid{0000-0003-4422-2919},
        S. Tavella\inst{1},
        N. Unger\inst{1}\orcid{0000-0003-3993-7127},
        G. Viviani\inst{2}\orcid{0009-0001-6201-2897}
}

\institute{Observatoire Astronomique de l’Université de Genève, Chemin Pegasi 51b, 1290 Versoix, Switzerland\\
  \email{emile.fontanet@unige.ch}         
           \and 
Institute of Physics, Laboratory of Astrophysics, Ecole Polytechnique Fédérale de Lausanne (EPFL), Observatoire de Sauverny, 1290 Versoix, Switzerland \and
Astronomy Unit, Queen Mary University of London, G.O. Jones Building, Bethnal Green, London E1 4NS, United Kingdom \and
Group of Applied Physics and Institute for Environmental Sciences, Université de Genève, Genève, Switzerland \and
Instituto de Astrofísica e Ciências do Espaço, Universidade do Porto,
CAUP, Rua das Estrelas, 4150-762 Porto, Portugal \and
Departamento de Física e Astronomia, Faculdade de Ciências,
Universidade do Porto, Rua do Campo Alegre, 4169-007 Porto,
Portugal 
}
   \date{Received 14 February 2025 / Accepted 11 April 2025
}
 
  \abstract
   {With around 200 detections of exoplanets around giant stars to date, our knowledge of the population of exoplanets orbiting evolved hosts more massive than the Sun remains limited. The CORALIE radial-velocity search for companions around evolved stars (CASCADES) was launched in 2006 with the aim of improving our understanding of the demographics of exoplanets around intermediate-mass stars, by studying them once they have evolved off the main sequence.}
   {We intend to refine the current sample of known exoplanets orbiting intermediate-mass (1.5 - 5 M$_\odot$) giant stars of spectral types G and early K. We searched for exoplanets orbiting the four stars HD 87816, HD 94890, HD 102888, and HD 121056.}
   {We used data obtained with the CORALIE spectrograph, mounted on the Leonhard Euler Swiss telescope located at La Silla Observatory in Chile. We gathered high-precision radial-velocity measurements over more than ten years for each of the aforementioned targets. We started by performing a search for periodic signals in the radial-velocity time series of the four targets by using periodograms. Following this, we fit for a Keplerian model using the significant peak with the highest power of the periodogram as the starting guess for the period. We then subtracted this model and repeated the procedure iteratively on the residuals until no significant peaks were found. Finally, to explore the posterior distribution of our models, the final solution was determined using a Markov chain Monte Carlo approach.}
   {We report the discovery of five new massive planets around HD 87816, HD 94890, and HD 102888 as well as the presence of a distant, potentially substellar, companion around HD 102888. We confirm the presence of a previously announced exoplanet orbiting the HD 121056 multi-object system with a period of 89 days and propose an update to the period of the outer companion.}
   {}

    \authorrunning{Fontanet et. al}
   \maketitle

\section{Introduction}
The number of exoplanet detections has risen from a few dozen 25 years ago to more than 5000 today. This rise in discoveries has provided the community with a wide range of new insights into planetary diversity. Planets are now found on all kinds of orbits, from close-in, hot terrestrial or Jupiter-sized planets to planets on orbits of thousands of days. Our knowledge of the demographics of exoplanets and their statistical properties has greatly increased in the past 30 years (\citealp{udry_statistical_2007}, \citealp{winn_occurrence_2015}, 
\citealp{zhu_exoplanet_2021}), and we have a much broader view today than we did 20 years ago. However, while the number of known exoplanets around stars with masses similar to or lower than that of the Sun has drastically grown, especially thanks to the transit method, detections around intermediate-mass stars, in the range 1.5-5 M$_\odot$, remain rather rare. 

Studying a wide range of stellar parameters, like mass and metallicity, for this population is crucial to improving our understanding of planet formation mechanisms. Depending on their mass, stars are expected to dissipate their disks at different rates (\citealp{kennedy_stellar_2009}), which could favor or hinder the formation of particular types of planets. For example, higher-mass stars are expected to disperse their disks more rapidly, limiting the duration of the planet formation process. The predicted time available for planet formation differs between the two main planet formation paradigms: core accretion (\citealp{pollack_formation_1996}, \citealp{alibert_models_2005}) and gravitational instability (\citealp{boss_giant_1997}, \citealp{durisen_gravitational_2007}). Testing these hypotheses with observational data is thus essential to improving our current models.

The CORALIE radial-velocity search for companions around evolved stars (CASCADES) survey was launched almost 20 years ago with the aim of filling this gap by focusing on intermediate-mass stars in their post-main-sequence phase (\citealp{ottoni_coralie_2022}). When on the main sequence, intermediate-mass stars (1.5-5 M$_\odot$, spectral type A to F) have high effective temperatures and rotational velocities, which result in a low number of spectral lines as well as a broadening of the lines. These characteristics represent a real challenge (see, e.g., \citealp{bouchy_fundamental_2001} on the impact of line broadening), limiting our ability to achieve the radial-velocity (RV) precision required to detect low-mass companions (\citealp{galland_extrasolar_2005-1}). When they evolve off the main sequence to the red giant branch, stars expand, which slows their rotation and decreases their surface temperature. This reduction in temperature and rotation rate facilitates more precise RV measurements, as a larger number of spectral lines become detectable via high-resolution spectroscopy and the effect of line broadening is reduced. Studying intermediate-mass stars in their evolved state, which we refer to here as giant stars, is thus a way to overcome the difficulties linked to them when they are still on the main sequence.\par
Additionally, studying the population of exoplanets around giant stars provides insights into the dynamical evolution of the planet-star connection. Planets with short orbital periods are expected to be engulfed by their host star during stellar evolution (\citealp{kunitomo_planet_2011}), which should also be reflected in the demographics through a lack of close-in planets. While hot Jupiters have been found around lower-mass evolved stars (M$_*$ < 1.5 M$_\odot$; \citealp{lillo-box_kepler-91b_2014}, \citealp{pereira_tess_2024}, \citealp{grunblatt_giant_2019}, \citealp{temmink_occurrence_2023}), they have yet to be found around more massive giant stars. Studying these targets is thus a way to understand if this phenomenon really happens, and how (\citealp{hon_close-giant_2023}, \citealp{lin_revealing_2024}).

\begin{figure}[h]
    \centering
    \includegraphics[width=\columnwidth]{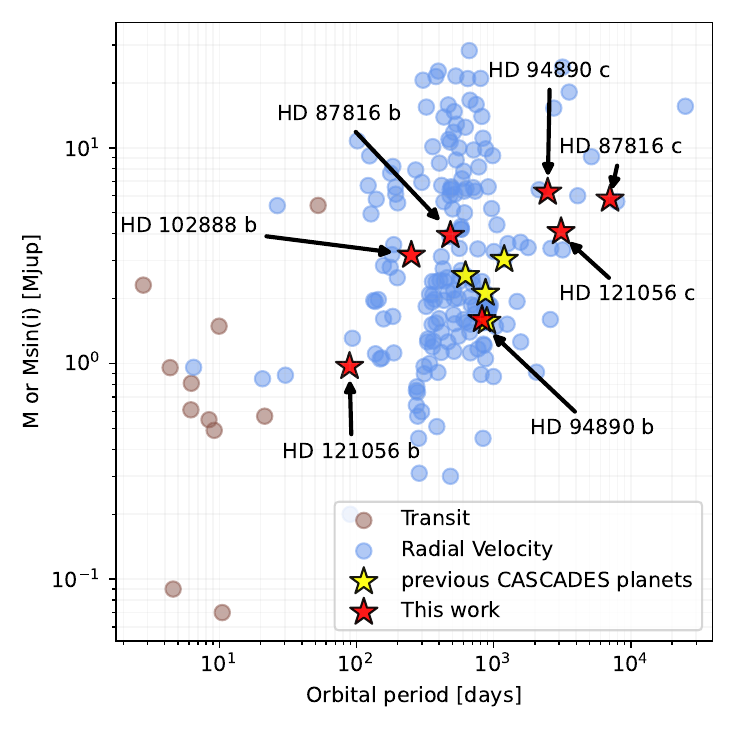}
    \caption{Exoplanet detections around giant stellar hosts (log $g$ < 3.5 cm$\,$s$^{-2}$) from the NASA Exoplanet archive. Circles are color-coded as a function of the detection method (blue for RV and brown for transit). Yellow stars represent the targets published by \citet{ottoni_coralie_2022} and \citet{pezzotti_coralie_2022}, and red stars the targets of this study.}
    \label{fig:detection_giant}
\end{figure}

Following the first detection of a planet in orbit around a giant star in 2002 by \citet{frink_discovery_2002}, approximately 200 exoplanets\footnote{This number was estimated with the help of the NASA Exoplanet Archive (\url{https://exoplanetarchive.ipac.caltech.edu}), by considering host stars with $\log g$ $\leq$ 3.5 cm\,s$^{-2}$.} have so far been detected in orbit around evolved hosts (as of January 2025; see Fig. \ref{fig:detection_giant}). As this number remains modest, our understanding of the demographics of this population of exoplanets can still be improved. Giant stars, being bright and easy to observe, act as excellent proxies for studying exoplanets around intermediate-mass stars. Consequently, several RV surveys have been conducted to study these stars (\citealp{frink_strategy_2001}, \citealp{sato_radial-velocity_2005}, \citealp{wittenmyer_pan-pacific_2011}, and \citealp{jones_study_2011}, among others). \par
However, studying the RV time series of giant stars can also be troublesome. While evolved stars offer certain observational advantages over their main-sequence counterparts, they are also known to be intrinsically variable. Signals of stellar origin, for instance caused by oscillations or convective motion, can produce periodic shifts in the RV time series of giant stars, which, if not accounted for, could be mistaken for a signal of planetary origin (\citealp{hekker_pulsations_2006}, \citealp{hekker_radial_2006}, \citealp{delgado_mena_planets_2023}). These activity-induced variations can occur over a wide range of periods, from several days to hundreds of days, with amplitudes exceeding 100 m$\,$s$^{-1}$. Disentangling true planetary signals from stellar activity is thus particularly challenging, requiring a careful analysis of the spectroscopic data to avoid false positives. 
 
In this paper we present the detection of five new exoplanets orbiting giant stars: two planets around HD 87816, two planets around HD 94890, and one planet around HD 102888. These targets have been monitored for over ten years as part of the CASCADES survey, allowing us to confirm their planetary nature and characterize their orbits. We also confirm the presence of the inner companion of HD 121056 with a period of 89 days, previously reported by \citet{feng_3d_2022}, \citet{wittenmyer_pan-pacific_2015} and \citet{jones_giant_2015}, and propose an update to the period and orbital parameters of HD 121056 c, for which previous values have varied across different studies. As we have access to a longer baseline of data than the previous studies of this star, we are confident that our revised period reflects the true orbital period of HD 121056 c. With fewer than 30 known multi-planet systems around giant stars, these discoveries provide valuable insights that help us understand the architecture of planetary systems in evolved stellar environments.\par
Section \ref{sec:cascades} provides a summary of the CASCADES survey, including our sample selection and the instrumental setup. In Sect. \ref{sec:obs} we present the observational data used for this work. Section \ref{sec:analysis} outlines our methods for planet detection and the process used to obtain their orbital parameters and posterior distributions; each system is also discussed individually. We conclude in Sect. \ref{sec:conclusion}.
\section{The CASCADES survey}
\label{sec:cascades}
\subsection{Sample description and history}
\begin{figure}[!htbp]
    \centering
    \includegraphics[width=\columnwidth]{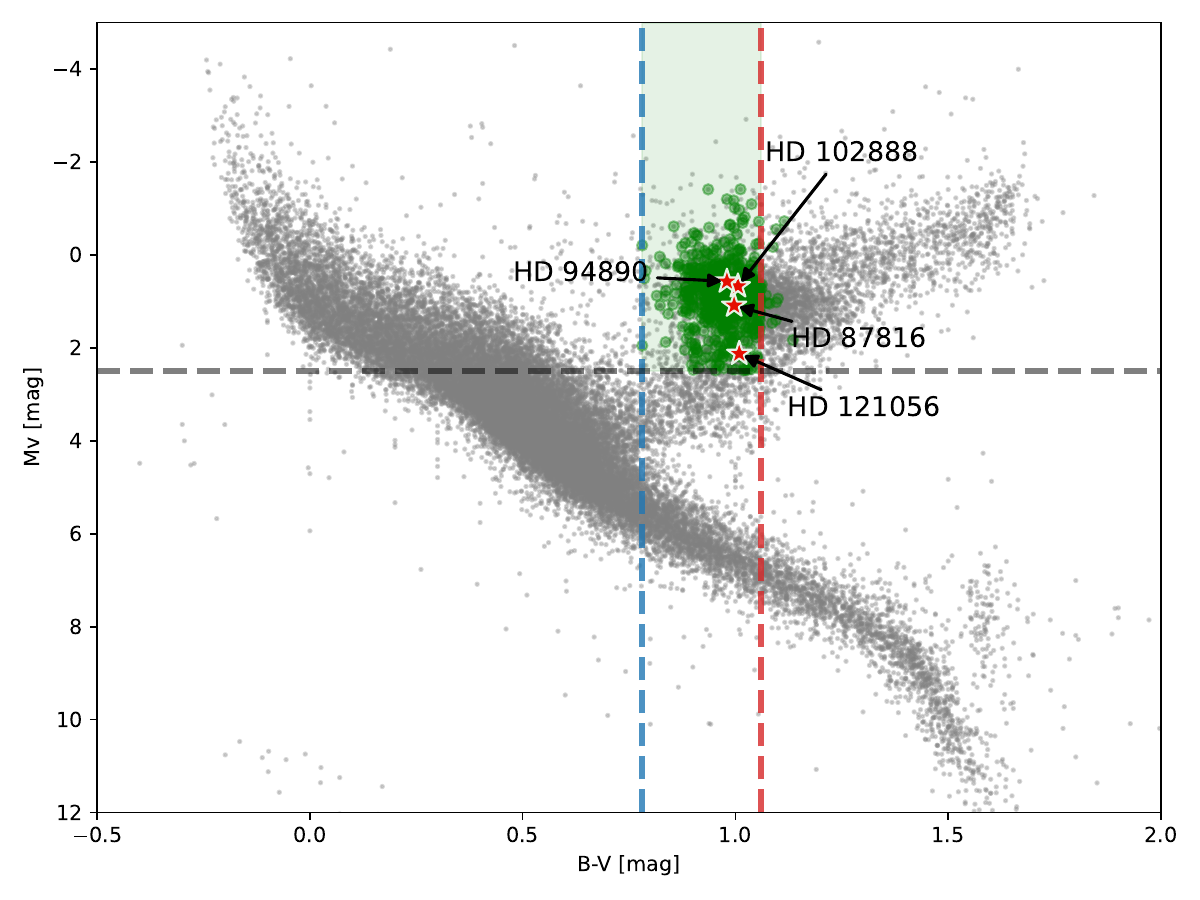}
    \caption{ CASCADES sample. Gray dots show all the stars of the \textsc{Hipparcos} sample with a precision on the parallax better than 14\%. The green circles show the 641 targets of the CASCADES survey. The blue and red vertical lines represent the two B-V cutoffs that were introduced to select the targets, and the horizontal gray line shows the absolute magnitude threshold that was used. The four targets that are discussed in this work are represented by the four red stars.}
    \label{fig:hr_diag}
\end{figure}
The CASCADES survey was launched in 2006 with the aim of performing a long-term, volume-limited survey of giant stars. It is conducted with the CORALIE spectrograph on the Leonhard Euler 1.2m Swiss telescope at La Silla, Chile. The 641 G- and K-type stars of the survey were selected from the \textsc{Hipparcos} catalog (\citealp{esa_hipparcos_1997}). Only stars located at less than 300 pc and with a precision on the parallax better than 14\% were considered (this limit was 10\% at first before the extension of the sample in 2011). Visual binaries (separated by less than 6") were discarded from the sample to prevent contamination by the secondary star. To isolate the giant stars in the Hertzsprung-Russell (HR) diagram, a M$_\text{V}$ cutoff was introduced, only keeping stars with M$_\text{V}\,$$ < 2.5$. Then, only stars with a B\,-\,V color index in the range 0.78\,< B\,-\,V <\,1.06 were kept, as can be seen in Fig. \ref{fig:hr_diag}. The magnitude upper limit and the color index lower bound ensure that stars have left the main sequence, while the higher bound of B-V was introduced to avoid later-type stars because of their higher activity level.\par

Estimates of stellar parameters, including masses and radii, were computed by \citet{ottoni_coralie_2022} for all of the 641 stars of the survey. Effective temperatures ($T_\text{eff}$), the surface gravity ($\log g$), and the metallicity ratio ([Fe/H]) were obtained from the analysis of high-resolution spectra. \citet{ottoni_coralie_2022} used the CORALIE spectra obtained after 2014 (after an instrument upgrade), which they stacked to create a high signal-to-noise master spectrum. The effective temperature $T_\text{eff}$, $\log g$ and [Fe/H] were obtained from these master spectra using the line list of \citet{tsantaki_deriving_2013} for the stars with $T_\text{eff}$ < 5200 K and that of \citet{sousa_spectroscopic_2008} for the others. They followed the method presented in \citet{sousa_ares_2014}, using the ARES (Automatic Routine for line Equivalent widths in stellar Spectra; \citealp{sousa_new_2007}, \citealp{sousa_ares_2015}) + MOOG (a widely used radiative transfer code for stellar abundance analysis; \citealp{sneden_carbon_1973}) methodology.\par
The luminosities were estimated by \citet{ottoni_coralie_2022} using the corrected parallaxes from \citet[obtained from the \textit{Gaia} DR2 parallaxes]{bailer-jones_estimating_2018}, the V-band magnitudes from \citet{hog_tycho-2_2000} and the bolometric correction relation of \citet{alonso_effective_1999}. Then, the radii were derived by using the Stefan-Boltzmann relation with the values of luminosity and effective temperatures obtained from the previous steps. Finally, the masses were estimated using the SPInS software (\citealp{lebreton_spins_2020}), using the position of the star in the HR diagram, $\log g$, and [Fe/H].\par
It is important to note that, for giant stars, it is hard to have a precise estimate of the mass from the evolutionary tracks. This is because, in this region of the HR diagram, evolutionary tracks of stars of different masses pile up in a very dense area. It is possible to obtain a better estimate of the masses by using more advanced techniques like asteroseismology, for instance. \citet{ottoni_coralie_2022} and \citet{buldgen_coralie_2022} computed the masses of four of the targets of the sample using asteroseismology and found that the masses are, on average, overestimated by the approach that uses SPInS. However, the masses obtained for the full sample by \citet{ottoni_coralie_2022} can still be used as a first reference for these stars. Fully detailed information on the survey can be found in \citet{ottoni_coralie_2022}, where it was originally presented.\par
As of January 2025, the CASCADES survey has detected four exoplanets around giant hosts (\citealp{ottoni_coralie_2022}, \citealp{pezzotti_coralie_2022}). More than 20 000 RV measurements have been made for the entire sample, with a median exposure time of 300 seconds and a mean precision of around 6 m$\,$s$^{-1}$. With more than 15 years of data, this survey is particularly well suited to identify long-period companions on periods of several thousand days.

\subsection{Instrument description}
The CORALIE spectrograph (\citealp{queloz_coralie_2000}) was installed in 1998 on the 1.2m Leonhard Euler Swiss telescope located at the La Silla observatory in Chile. CORALIE is a two-fiber-fed echelle spectrograph working in the visible (3880-6810 \r{A}). The instrument received two major hardware upgrades in 2007 and in 2014, which resulted in increases in overall throughput and resolution (see, e.g., \citealp{segransan_coralie_2010}). With these updates, the instrument's optomechanical configuration changed, which led us to treat data coming from these different periods as if they came from different instruments. For this reason, we refer to the pre-upgrade data as CORALIE98, the ones after the 2007 upgrade as CORALIE07, and the latest ones as CORALIE14. For the analysis, we thus used different instrument offsets and precision for these three versions of CORALIE. From the analysis of quiet stars in the survey, which we consider to be constant, we have determined that the instrumental stability for each version of the instrument is typically $\sigma_\text{COR98}$ = 5 m$\,$s$^{-1}$, $\sigma_\text{COR07}$ = 8 m$\,$s$^{-1}$ and $\sigma_\text{COR14}$ = 3 m$\,$s$^{-1}$. Temperature and air pressure variations during the night can lead to a spectral drift, which induces systematic shifts in the wavelength calibration of the instrument. If not accounted for, the drift can introduce errors in the RV measurements. As CORALIE is not kept in a vacuum vessel, changes in the weather conditions typically introduce a drift that is tracked throughout the night. When the drift exceeds 110 m$\,$s$^{-1}$, the CORALIE data reduction software can no longer reliably correct for it, and the quality control (QC) flag is set to "Failed."\par
\begin{table*}[h]

\raggedright

\caption{Stellar parameters for HD 87816, HD 94890, HD 102888, and HD 121056.}
    \centering
\begin{tabular}{llccccc}
\toprule
\toprule
Quantity&Unit&Ref. &                     HD 87816 &                     HD 94890 &                    HD 102888 &                    HD 121056 \\
\midrule
   Spectral type & &[1, 5] &                     K0III &                     K0III &                     G8III &                     K0III \\
       $V$ &[mag] &[2, 5] &     6.50 ± 0.01 &     4.60 ± 0.001 &     6.48 ± 0.01 &     6.19 ± 0.01 \\
       
     $B-V$& [mag]&[2, 5] &       0.99 ± 0.015 &     1.0 ± 0.01 &      0.96 ± 0.02&      1.0 ± 0.02 \\
      $\pi$ &[mas] & [3, 5] &           7.48 ± 0.03 &          16.18 ± 0.19 &           7.88 ± 0.07 &          15.57 ± 0.04 \\
      $M_\text{v}$ &[mag] & [5]&  0.88 ± 0.01& 0.65 ± 0.03 & 0.97 ± 0.02 &  2.15 ± 0.01 \\
       
       $d$ & [pc]& [4, 5]& 133.23 ± 0.541 & 61.70 ± 0.71 & 126.48 ± 1.14 & 64.12 ± 0.17 \\
   $T\textsubscript{eff}$ &[K] &[3, 5] &             4989 ± 46 &             4867 ± 38 &             4990 ± 39&             4840 ± 43 \\
   $\log g$ &[cm\,s$^{-2}$] &[5] &              2.86 ± 0.1 &               2.6 ± 0.09 &              2.87 ± 0.09 &              3.11 ± 0.13 \\
    {[Fe/H]} & [dex]& [5]&             0.14 ± 0.04 &             -0.02 ± 0.03 &             0.07 ± 0.03 &             0.01 ± 0.03 \\
   $M_*$ & [M$_\odot$]& [5]& 2.41 ± 0.10 & 1.74 ± 0.196 & 2.42 ± 0.06 & 1.28 ± 0.08 \\
   $L_*$ & [L$_\odot$]&[5] & 45.2 ± 0.93 & 58.01 ± 1.64 & 41.29 ± 1.04 & 14.68 ± 0.27 \\
  $R_*$ & [R$_\odot$] &[5] & 9.0 ± 0.19 & 10.71 ± 0.23 & 8.6 ± 0.17 & 5.45 ± 0.11 \\
\bottomrule
\end{tabular}\\
\vspace{1em}
\raggedright
\tablebib{[1]~\textsc{Hipparcos} catalog \citet{esa_hipparcos_1997}; [2]~Tycho-2 catalog (\citealp{hog_tycho-2_2000}); [3]~\textit{Gaia} DR2 (\citealp{gaia_collaboration_gaia_2018}); [4]~\citet{bailer-jones_estimating_2018}; [5]~\citet{ottoni_coralie_2022}}
\label{tab:star_params}
\end{table*}
One of the targets of this work, HD 121056, was also observed with the FEROS (\citealp{kaufer_commissioning_1999}) and CHIRON (\citealp{schwab_performance_2012} and \citealp{spronck_chiron_2011}) spectrographs. We present here a combined solution including CORALIE, CHIRON, and FEROS, based on a longer time span of observations.

\section{Observations and stellar characteristics}

\label{sec:obs}

All the stellar parameters of the four stars discussed in this work can be found in Table \ref{tab:star_params}. A summary of the observational data available for all targets is presented in Table \ref{tab:obs_summary}. The CORALIE data reduction software versions used for this work are 3.3 for CORALIE98, 3.4 for CORALIE07, and 3.8 for CORALIE14.

\begin{table}
\caption{Summary of observations.}             
\label{tab:obs_summary}      
\centering                          
\begin{tabular}{c c c c }        
\toprule\toprule               
HD & N$_\text{meas}$ & med($\sigma_\text{RV}$) [m$\,$s$^{-1}$] & T$_\text{span}$ [yr] \\    
\midrule   
   87816 & 91 & 4.7 & 18 \\      
   94890 & 76 & 2.6    & 18.1 \\
   102888 & 36 & 3.5     & 12.5 \\
   121056 & 105 & 4.6    & 16.6 \\
\bottomrule                                   
\end{tabular}
\end{table}

\section{Analysis of the radial velocity time series}
\label{sec:analysis}
The monitoring of the targets of the sample and the identification of possible planetary signals were conducted using the Data \& Analysis Center for Exoplanets (DACE) web tool\footnote{\url{https://dace.unige.ch/dashboard/}}, which allows us to visualize the time series of RVs and line profile indicators, as well as their periodograms (\citealp{zechmeister_generalised_2009}). A more in-depth analysis of the RV time series was then performed using the Kepmodel python package (\citealp{delisle_analytical_2022}).\par
We only considered the observations that passed the CORALIE QC, and discarded all measurements with an uncertainty on the RV exceeding 20 m$\,$s$^{-1}$ (which were obtained in poor weather conditions). We corrected all the RV time series for secular acceleration using the latest values of parallax and proper motion in right ascension and declination available from the SIMBAD database (\citealp{wenger_simbad_2000}), based on \textit{Gaia} Early Data Release 3 (\citealp{riello_gaia_2021}). Although included for completeness, this effect is negligible compared to the intrinsic variability of the stars at these distances. Then, we fit an RV offset for each dataset, treating different versions of the CORALIE spectrograph as independent instruments. We also included in the fit our assumptions on the stability of the different versions of CORALIE (5 m$\,$s$^{-1}$, 8 m$\,$s$^{-1}$, and 3 m$\,$s$^{-1}$ for CORALIE98, 07, and 14, respectively). To account for additional noise from stellar activity-related effects, we also included a global stellar jitter term.\par
To identify significant periodic signals, we conducted an iterative periodogram analysis. A Keplerian model was fit to the dominant peak in the periodogram if its false alarm probability (FAP; using that provided by Kepmodel) was below $10^{-2}$. This process was repeated iteratively on the residuals until no significant peak could be detected. Each detected signal was subsequently refined using a full Keplerian fit. As highlighted in studies of stable giant stars by \citet{frink_strategy_2001} and \citet{hekker_precise_2006}, the intrinsic RV scatter for quiet giant stars can typically reach 20 m$\,$s$^{-1}$. Consequently, we consider any residual scatter of this magnitude to be consistent with the expected stellar jitter rather than indicative of an additional astrophysical signal.\par
The final orbital parameters for each system, presented in Tables \ref{tab:87816_params} to \ref{tab:hd121056_params}, were derived using the Markov chain Monte Carlo (MCMC) algorithm implemented in SAMSAM (\citealp{delisle_samsam_2022}). The MCMC was run for 300 000 iterations and modeled the following orbital parameters: orbital period ($P)$, RV semi-amplitude ($K$), mean longitude ($\lambda_0$), and the parameters $\sqrt{e} \cdot \cos{\omega}$ and $\sqrt{e} \cdot \sin{\omega}$, where $e$ and $\omega$ are the orbital eccentricity and the argument of periastron, respectively. The different offsets $\gamma_i$, and instrumental precisions $\sigma_i$ (for CHIRON and FEROS), where $i$ corresponds to the instrument name, were also sampled with the MCMC.

The mean longitude, $\sqrt{e} \cdot \cos{\omega}$ and $\sqrt{e} \cdot \sin{\omega}$ are sampled with uniform priors, while the orbital period and the RV amplitude are sampled with a log-uniform prior. We also used a uniform prior for the COR14 offset of reference and for the relative offsets between COR98/COR07 and COR14: $\Delta \gamma_\text{COR98/07 - COR14}$ = $\mathcal{U}(-50, 50)$, as larger offset values between the different versions of the instrument are not expected. The stellar jitter term $\sigma_\text{jit}$ and the CHIRON and FEROS instrument stability are sampled with uniform priors. The full list of priors used for the MCMC can be found in Table \ref{tab:priors}.

\begin{table}[h]
    \centering
    \caption{Priors used for the MCMC sampling.}
    \begin{tabular}{ccccc}
    \toprule
    \toprule
       Parameter  & Prior & LB & UB&Unit\\
       \midrule
        $P$ & log$\mathcal{U}$ & 5 & 5$\cdot10^4$& [d]\\
        $K$ & log$\mathcal{U}$ & 0.1 & 10$^4$ & [m$\,$s$^{-1}$]\\
        $\lambda_0$ & $\mathcal{U}$& 0&2$\pi$ &[rad]\\
        $\sqrt{e} \cdot \sin{\omega}$ &$\mathcal{U}$& -1 & 1 \\
        $\sqrt{e} \cdot \cos{\omega}$ & $\mathcal{U}$& -1 & 1 \\
        $\gamma_\text{COR98-COR14}$ & $\mathcal{U}$& -50 &50  &[m$\,$s$^{-1}$] \\
        $\gamma_\text{COR07-COR14}$ & $\mathcal{U}$& -50 &50 &[m$\,$s$^{-1}$]  \\
        $\gamma_\text{COR14}$ & $\mathcal{U}$ & RV$_\text{min}$ & RV$_\text{max}$&[m$\,$s$^{-1}$]\\
        $\sigma_\text{jit}$ & $\mathcal{U}$&0 &30 &[m$\,$s$^{-1}$]\\

        $\sigma_\text{CHIRON}$ & $\mathcal{U}$&  0& 10&[m$\,$s$^{-1}$]\\
        $\sigma_\text{FEROS}$ & $\mathcal{U}$& 0 & 10&[m$\,$s$^{-1}$]\\
        $\gamma_\text{CHIRON}$ & $\mathcal{U}$ & RV$_\text{min}$ & RV$_\text{max}$ &[m$\,$s$^{-1}$]\\
        $\gamma_\text{FEROS}$ & $\mathcal{U}$ & RV$_\text{min}$ & RV$_\text{max}$&[m$\,$s$^{-1}$]\\
         \bottomrule
    \end{tabular}
    
    \label{tab:priors}
    \tablefoot{LB and UB refer to the lower and upper bounds of the prior distributions, respectively.}

\end{table}

\subsection{The HD 87816 planetary system: An eccentric inner planet and a massive, long-period outer one}

\begin{figure}[h]
    \centering
    \includegraphics[width=\columnwidth]{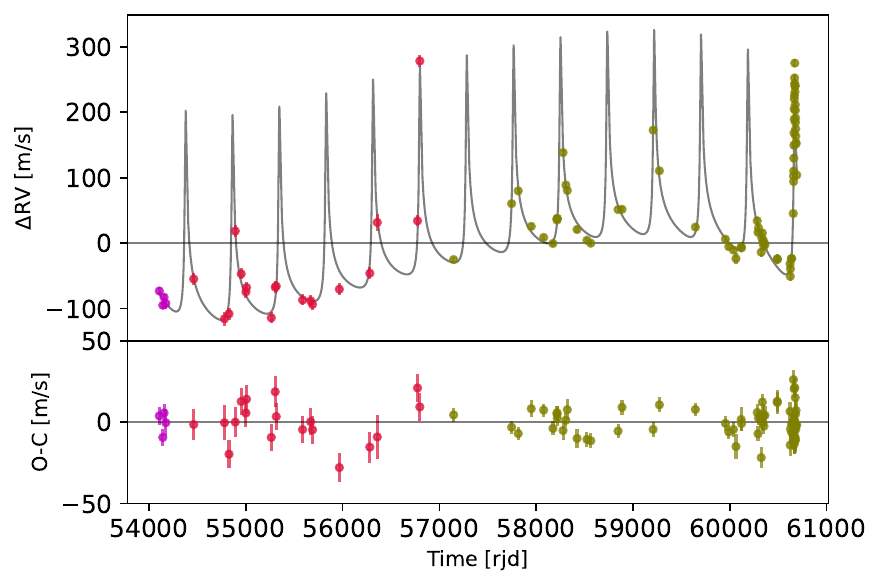}
    \caption{Top: RV time series of HD 87816 obtained between 2007 and 2024. The dots are color-coded by instrument (CORALIE98 in purple, CORALIE07 in red, and CORALIE14 in olive). Our two-planet model is overplotted in black, with one very eccentric planet orbiting with a period of 484 days and a longer-period planet at 7600 days. Bottom: Residuals after subtracting the two-planet model.}
    \label{fig:87816_timeseries}
\end{figure}

\begin{figure}[h]
    \centering
    \includegraphics[width=\columnwidth]{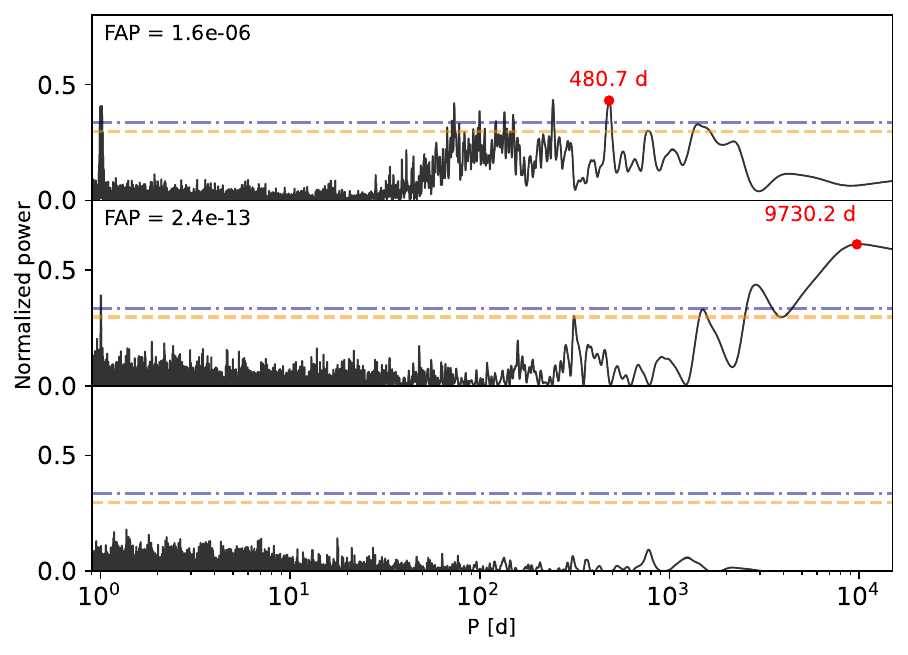}
    \caption{Top: Periodogram of the RV time series of HD 87816. Middle: Periodogram obtained after fitting a first Keplerian model with a period of 480 days. Bottom: Periodogram of the residuals of the RV time series after subtraction of the two-planet model. The horizontal lines show the different FAP levels: 1\% (dashed orange line) and 0.1\% (dashed-dotted blue line). The highest peak is marked with a red dot, and the corresponding period is indicated next to it.}
    \label{fig:87816_periodo_planet}
\end{figure}

\begin{figure}[h]
    \centering
    \includegraphics[width=\columnwidth]{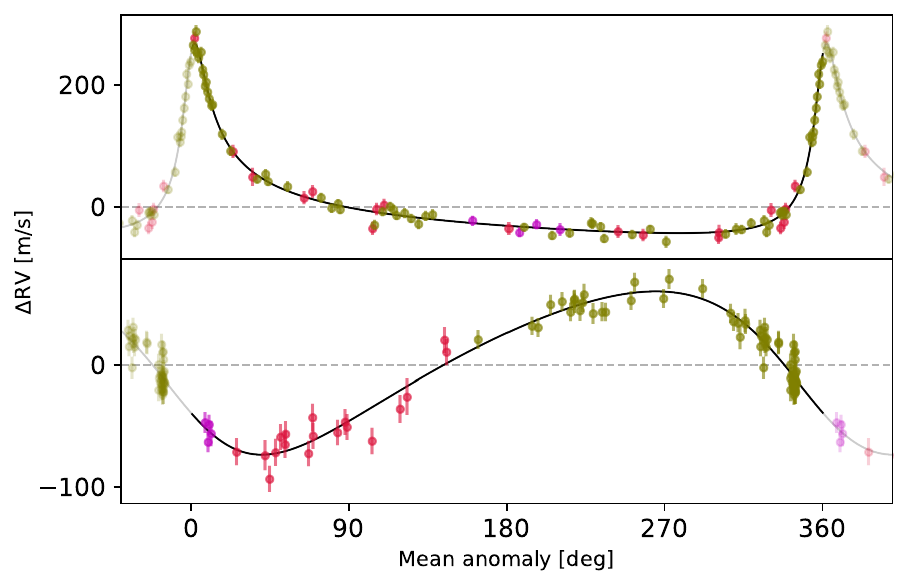}
    \caption{Top: Phase-folded RVs for the best model obtained for HD 87816 b. Bottom: Same but for HD 87816 c.}
    \label{fig:87816_phasefold}
\end{figure}

HD 87816 has been observed with CORALIE for approximately 18 years, starting in 2007, with a total of 91 individual observations. Via analysis of its RV time series, we identified two candidates (Figs. \ref{fig:87816_timeseries} to \ref{fig:87816_phasefold}), one of which is on a very eccentric orbit.\par
The periodograms obtained from the RV time series can be seen in Fig. \ref{fig:87816_periodo_planet}. After fitting for the two Keplerian orbits, one can see that the residuals do not show any remaining significant periodic signal. The full orbital parameters obtained from the MCMC analysis can be found in Table \ref{tab:87816_params}.\par
HD 87816 b, a planet with a minimum mass $m_\text{b}\sin i_\text{b}$ = 6.7 M$_\text{Jup}$, takes 484 days to orbit its host star on an orbit with an eccentricity of $e$ = 0.78, making it one of the most eccentric RV-detected planets around a giant star\footnote{As of January 2025, only KIC 3526061 b ($e$ = 0.85, \citealp{karjalainen_companions_2022}) and HD 76920 b ($e$ = 0.86, \citealp{wittenmyer_pan-pacific_2017}) are known to be on more eccentric orbits.}. As the periastron passage of HD 87816 b was expected to occur in December 2024, we increased our observing cadence of this star during this period to obtain a good sampling of the rapid rise and fall of the RVs at this passage of the orbit. This can be observed in Fig. \ref{fig:87816_timeseries}.\par
HD 87816 c, the second planet of the system with a minimum mass of $m_\text{c}\sin i_\text{c}$ = 12.2 M$_\text{Jup}$, orbits its host star with a period of approximately 7600 days and an eccentricity of $e$ = 0.19. As the period of HD 87816 c exceeds the time span of our observations, its value cannot be easily constrained from our analysis. This difficulty arises from the combination of the long orbital period and the need to account for different offset values associated with the three versions of CORALIE. This can be seen in Fig. \ref{fig:corner_87816}, where clear correlations between the period of HD 87816 c and the instrumental offsets can be seen. Consequently, this explains the large uncertainties that we obtain from our MCMC analysis. We also note that the imprecision of mass estimation models for giant stars adds to the uncertainty on the mass of the outer companion.\par

\begin{table}[h!]
\centering
\caption{Orbital parameters obtained for HD 87816 b and c.}
\begin{tabular}{lccc}
\toprule
\toprule

Parameter&Unit &  HD 87816 b & HD 87816 c\\
\midrule
$P$&[days]  & $484.17^{+0.13}_{-0.12}$ &  $7596^{+1140}_{-535}$ \\[3pt]
$K$&[m$\,$s$^{-1}$]  & $155.03^{+2.80}_{-2.78}$  &  $71.30^{+9.56}_{-7.59}$ \\[3pt]
$e$&  &  $0.78^{+0.005}_{-0.005}$ & $0.19^{+0.07}_{-0.07}$  \\[3pt]
$\omega$&[deg]  & $337.10^{+1.18}_{-1.19}$  & $110.44^{+16.43}_{-22.61}$ \\[3pt]
$\lambda_0$&[deg]  & $8.68^{+0.92}_{-0.94}$ &  $5.83^{+5.59}_{-4.46}$ \\[3pt]
$a$&[au]  & $1.618^{+0.0003}_{-0.0003}$ &  $10.14^{+0.99}_{-0.48}$ \\[3pt]
$m_\text{p}$ sin $i$ & [M$_\text{Jup}$] &  $6.74^{+0.13}_{-0.13}$ & $12.20^{+2.15}_{-1.59}$ \\[3pt]
\midrule
$\gamma_\text{COR98}$&[m$\,$s$^{-1}$] & \multicolumn{2}{c}{$5585.33^{+11.81}_{-13.59}$} \\[3pt]
$\gamma_\text{COR07}$ &[m$\,$s$^{-1}$] & \multicolumn{2}{c}{$5596.54^{+9.96}_{-15.87}$}  \\[3pt]
$\gamma_\text{COR14}$&[m$\,$s$^{-1}$] & \multicolumn{2}{c}{$5596.41^{+8.28}_{-11.72}$}  \\[3pt]
\midrule
$\sigma_\text{jit}$ &[m$\,$s$^{-1}$]  & \multicolumn{2}{c}{$7.65^{+1.06}_{-1.01}$}  \\[3pt]
\bottomrule
\end{tabular}
\label{tab:87816_params}
\tablefoot{The value of $\lambda_0$ is given for JD = 2458775.29}
\end{table}

Periodograms of the time series of the full width at half maximum (FWHM) of the cross-correlation function, bisector inverse slope (BIS), and the H-alpha can be seen in Fig. \ref{fig:87816_activity}. Following recent observations to complete the orbital coverage of HD 87816 b, a significant peak emerged in the periodogram of the FWHM time series at 519 days. We checked for a correlation between the FWHM and the RVs but did not find any. Thanks to the long baseline of the observations, the two peaks, corresponding respectively to the 484-day orbital period of HD 87816 b and the 519-day signal for the FWHM variation, are clearly separated in the frequency domain. This is supporting our confidence in the planetary nature of the 484-day signal. However, the origin of the 519-day signal still remains unclear. A possible explanation could be that it results from a long-term variation in the FWHM time series (potentially due to, e.g., a long-period double-lined spectroscopic binary or a long-timescale magnetic cycle) convolved with our sampling window. This interpretation is supported by the fact that the signal completely vanishes if we discard the last 26 high-cadence observations. No significant peak can be seen for any of the other activity-index time series, reinforcing the interpretation of a planetary origin of the periodic signals observed in the RVs.

\subsection{Two massive planets around HD 94890}
We have 76 RV measurements of HD 94890, conducted with the three different versions of CORALIE. This star was observed between 2007 and 2025, and we report here the detection of a two-planet system around it.\par
\begin{figure}[h]
    \centering
    \includegraphics[width=0.935\columnwidth]{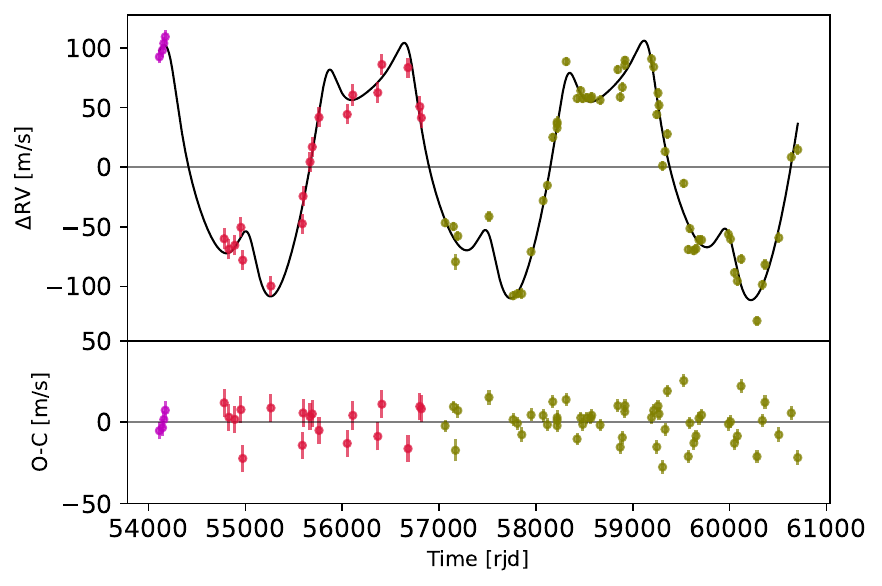}
    \caption{Same as Fig. \ref{fig:87816_timeseries} but for HD 94890.}
    \label{fig:94890_rvtseries}
\end{figure}

\begin{figure}[h]
    \centering
    \includegraphics[width=0.91\columnwidth]{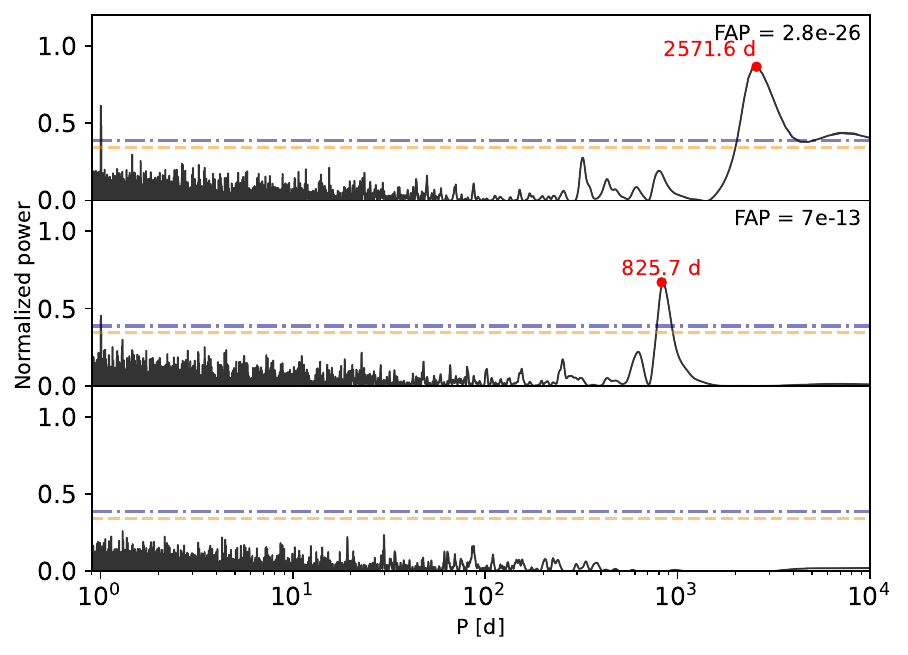}
    \caption{Same as Fig. \ref{fig:87816_periodo_planet} but for HD 94890.}
    \label{fig:94890_periodo}
\end{figure}
\begin{figure}[h]
    \centering
    \includegraphics[width=\columnwidth]{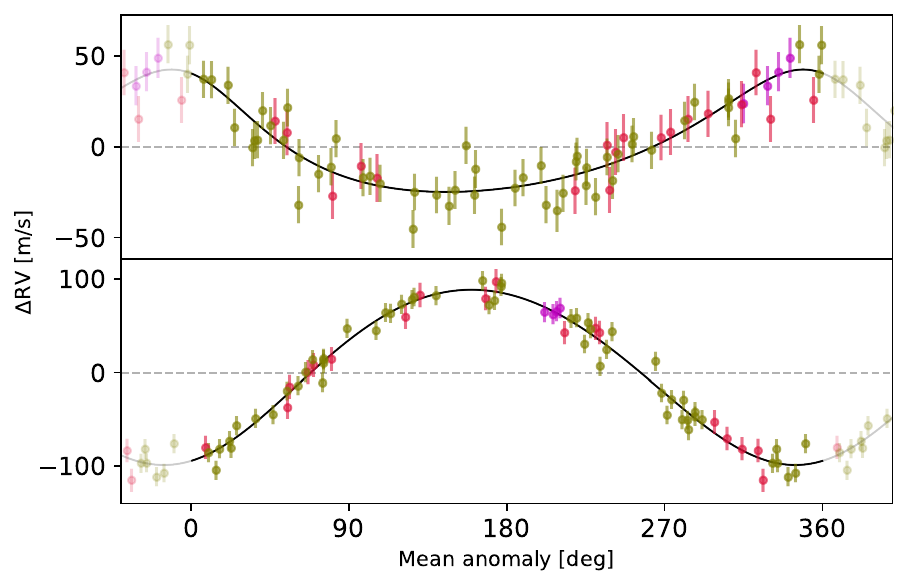}
    \caption{Top: Phase-folded orbit of HD 94890 b obtained with CORALIE. Bottom: Same but for HD 94890 c.}
    \label{fig:94890_phasefold}
\end{figure}
\begin{table}[h!]
\centering
\caption{Orbital parameters obtained for HD 94890 b and c.}
\begin{tabular}{lccc}
\toprule
\toprule
Parameter&Unit &HD 94890 b & HD 94890 c \\
\midrule
$P$&[days]  & $824.61^{+5.02}_{-4.80}$ & $2492.19^{+14.72}_{-13.94}$ \\[3pt]
$K$&[m$\,$s$^{-1}$]  & $32.85^{+2.69}_{-2.66}$ & $92.43^{+2.54}_{-2.62}$ \\[3pt]
$e$&  & $0.22^{+0.08}_{-0.09}$ & $0.05^{+0.03}_{-0.03}$ \\[3pt]
$\omega$&[deg]  & $19.68^{+17.05}_{-18.81}$ & $200.23^{+26.17}_{-26.96}$\\[3pt]
$\lambda_0$&[deg]  & $205.73^{+3.97}_{-4.03}$ & $232.65^{+1.44}_{-1.40}$ \\[3pt]
$a$&[au]  & $2.07^{+0.01}_{-0.01}$ & $4.33^{+0.02}_{-0.02}$ \\[3pt]
$m_\text{p}$ sin $i$ & [M$_\text{Jup}$] & $2.13^{+0.16}_{-0.17}$ & $8.91^{+0.24}_{-0.25}$ \\[3pt]
\midrule
$\gamma_\text{COR98}$&[m$\,$s$^{-1}$] & \multicolumn{2}{c}{$2170.31^{+10.08}_{-9.80}$} \\[3pt]
$\gamma_\text{COR07}$ &[m$\,$s$^{-1}$] & \multicolumn{2}{c}{$2179.17^{+3.76}_{-3.73}$}  \\[3pt]
$\gamma_\text{COR14}$&[m$\,$s$^{-1}$] & \multicolumn{2}{c}{$2189.52^{+1.71}_{-1.74}$}  \\[3pt]
\midrule

$\sigma_\text{jit}$ &[m$\,$s$^{-1}$]  & \multicolumn{2}{c}{$10.92^{+1.28}_{-1.12}$}  \\[3pt]

\bottomrule
\end{tabular}
\label{tab:hd94890_params}
\tablefoot{The value of $\lambda_0$ is given for JD = 2457939.04}
\end{table}

The RV time series of HD 94890 can be seen in Fig. \ref{fig:94890_rvtseries}, where a clear pattern is noticeable without the need for an advanced analysis. We followed the same periodogram approach as we did for the other stars in this work. All the periodograms of each step of the iterative fitting can be found in Fig. \ref{fig:94890_periodo}. Assuming a stellar mass $M_*$ = 1.74 M$_\odot$ for HD 94890, we find planetary masses of 2.1 M$_\text{Jup}$ and 8.9 M$_\text{Jup}$ for HD 94890 b and c, respectively. Both planets are on relatively low-eccentricity orbits, with $e$ = 0.22 for HD 94890 b and $e$ = 0.05 for HD 94890 c. After the subtraction of our best two-planet model, no significant peak remains in the periodogram of the RV time series. The residuals (Fig. \ref{fig:94890_rvtseries}, bottom) do not show any clear periodicity and seem compatible with the RV scatter expected for giant stars. HD 94890 b and HD 94890 c orbit their host star with periods of 824 days and 2492 days, respectively, resulting in a period ratio notably close to 3. This value suggests a potential 3:1 mean-motion resonance, wherein the inner planet completes three orbits for every orbit of the outer one. Such a resonant configuration can lead to significant gravitational interactions between the planets, potentially affecting the observed RV of the host star and making it differ from that expected for two noninteracting Keplerian orbits. Moreover, these interactions depend on the true masses and orbital inclinations of the two planets in resonance. A detailed dynamical analysis could therefore constrain the orbital inclination and yield more accurate estimates of the planetary masses. While such an investigation is beyond the scope of this work, it represents a promising direction for future research and could significantly enhance our understanding of this system.\par
The periodograms of the time series of FWHM, H-alpha, and BIS (Fig. \ref{fig:94890periodo_activity}) do not show any significant peak. We also checked for correlations between these quantities and the RVs, without finding any. The phase-folded RVs of HD 94890 b and HD 94890 c can be found in Fig. \ref{fig:94890_phasefold}. Our data fully cover both orbits, and thus we do not need additional points to fill missing phases.

\subsection{A planet and an unseen companion around HD 102888}
\begin{figure}[h]
    \centering
    \includegraphics[width=\columnwidth]{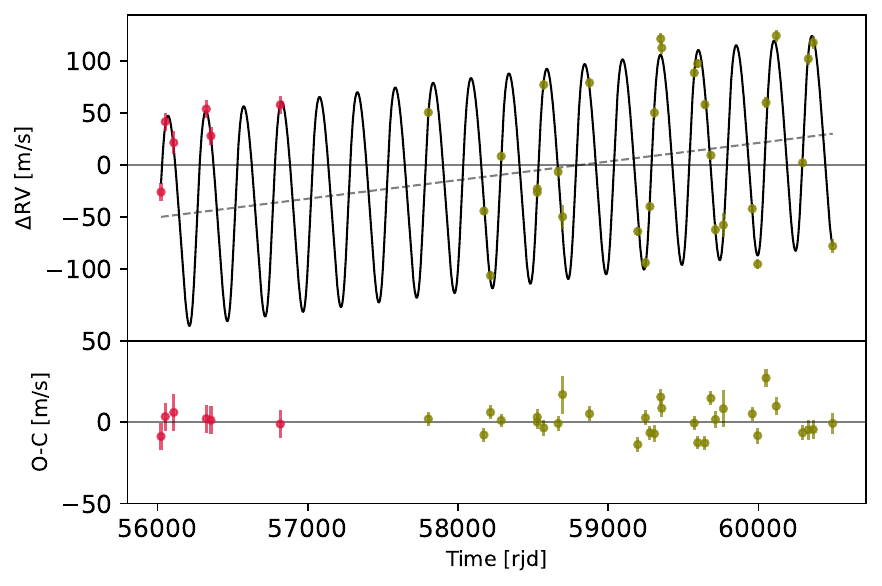}
    \caption{Same as Fig. \ref{fig:87816_timeseries} but for HD 102888.}
    \label{fig:102888_rvtseries}
\end{figure}

HD 102888 has been observed for 12 years with the CORALIE spectrograph between 2012 and 2024. By looking at the 36 CORALIE RV measurements from Fig. \ref{fig:102888_rvtseries}, a 200 m$\,$s$^{-1}$ peak-to-peak periodic variation of the RV can be seen quite easily.\par

\begin{figure}[h]
    \centering
    \includegraphics[width=\columnwidth]{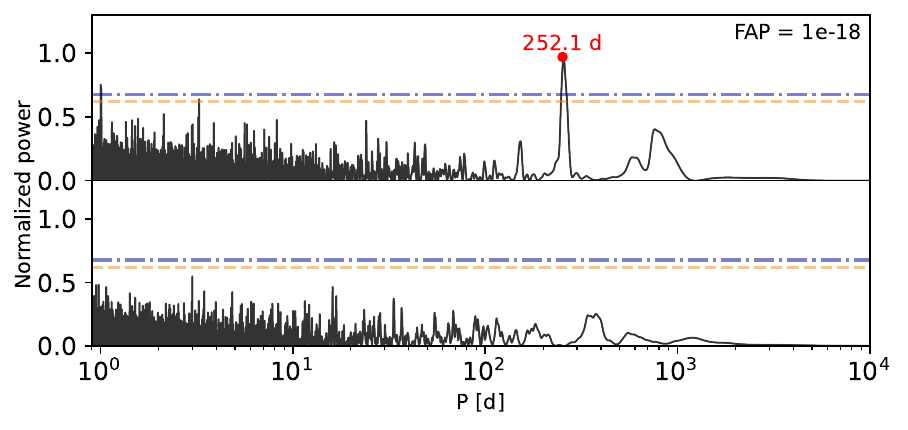}
    \caption{Same as Fig. \ref{fig:87816_periodo_planet} but for HD 102888 after subtracting the linear drift.}
    \label{fig:102888_rv_periodogram}
\end{figure}
\begin{figure}[h]
    \centering
    \includegraphics[width=\columnwidth]{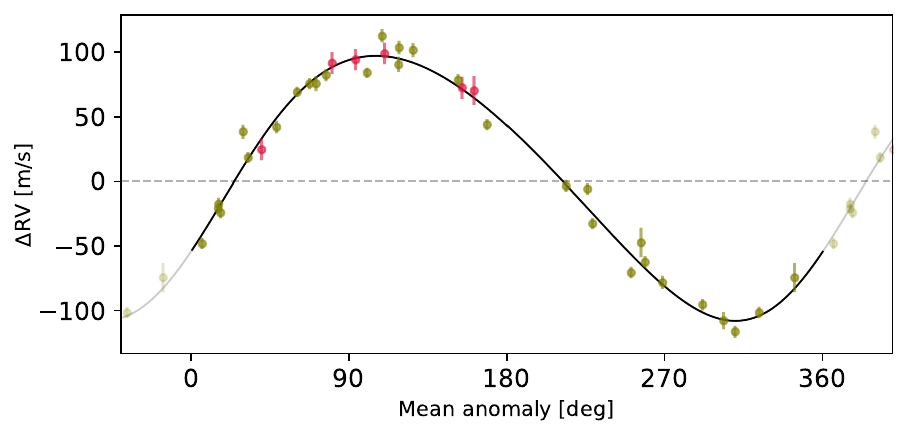}
    \caption{Phase-folded RV data of HD 102888 with the period of HD 102888 b at 252 days. Overplotted in black is the best-fit Keplerian model.}
    \label{fig:102888_phasefold}
\end{figure}
\begin{table}[h!]

\centering
\caption{Orbital parameters obtained for the system around HD 102888.}
\begin{tabular}{lcc}
\toprule
\toprule
Parameter&Unit & HD 102888 b\\
\midrule

$P$&[days]   & $252.06^{+0.24}_{-0.24}$ \\[3pt]
$K$&[m$\,$s$^{-1}$]  & $102.30^{+2.79}_{-2.80}$  \\[3pt]
$e$&   & $0.11^{+0.03}_{-0.03}$\\[3pt]
$\omega$&[deg]   & $241.20^{+13.33}_{-13.62}$\\[3pt]
$\lambda_0$&[deg]   & $307.39^{+1.55}_{-1.52}$ \\[3pt]
$a$&[au]  & $1.05^{+0.001}_{-0.001}$\\[3pt]
$m_\text{p}$ sin $i$ & [M$_\text{Jup}$]  & $5.69^{+0.15}_{-0.15}$ \\[3pt]
\midrule
$\gamma_\text{COR07}$ &[m$\,$s$^{-1}$] & $-2238.13^{+8.87}_{-8.92}$  \\[3pt]
$\gamma_\text{COR14}$&[m$\,$s$^{-1}$] & $-2230.15^{+2.40}_{-2.38}$ \\[3pt]
$\dot{\gamma}$ & [m$\,$s$^{-1}$yr$^{-1}$] & $6.42^{+1.00}_{-1.02}$  \\[3pt]

\midrule

$\sigma_\text{jit}$ &[m$\,$s$^{-1}$]  & $8.66^{+1.81}_{-1.49}$  \\[3pt]

\bottomrule
\end{tabular}
\label{tab:hd102888}
\tablefoot{The value of $\lambda_0$ is given for JD = 2458815.21}
\end{table}

After fitting a Keplerian model to this periodic signal, we noticed a remaining long-term trend in the RV data, which we modeled with a linear drift. Both the drift and the model for the Keplerian orbit can be seen alongside the data in Fig. \ref{fig:102888_rvtseries}. The period of the variation was identified using a periodogram approach. The periodogram of the RV time series can be found in Fig. \ref{fig:102888_rv_periodogram}. The peak at 252 days is easily identifiable and well above the FAP threshold for significant detection (0.01). After fitting for the planet and the drift, no significant peak can be seen on the periodogram of the residuals. A phase-folding of the data with the best Keplerian model is shown in Fig. \ref{fig:102888_phasefold}. Knowing that activity-induced RV variations could be at the origin of periodic signals, we looked for hints of stellar activity-related periodic signals in the FWHM, BIS, and H-alpha index time series of HD 102888. The resulting periodograms can be found in Fig. \ref{fig:102888_activityl}, where we see that no periodic variation of any of these quantities is found.\par

From our analysis, HD 102888 b is a planet with minimum mass of $m_\text{b}\sin i_\text{b}$ = 5.7 M$_\text{Jup}$ orbiting its star in 252 days on an orbit with an amplitude $K$ = 102 m$\,$s$^{-1}$ and eccentricity of $e$ = 0.1. The remaining linear trend after subtraction of the Keplerian model from our data corresponds to an acceleration of $\dot{\gamma}$ = 6.4 $\pm$ 1 m$\,$s$^{-1}$yr$^{-1}$. This hints at the presence of an additional unseen companion on a much larger orbit that far exceeds our observation baseline. Following \citet{winn_hat-p-13_2010}, and assuming that this companion is on a circular orbit and that its mass is much smaller than that of HD 102888, we can approximate $\dot{\gamma}$ $\sim$ $GM_\text{c}\sin i_\text{c}/a^2_\text{c}$ to have an estimate on the order of magnitude of the mass of this companion. In our case, this yields
\begin{equation}
\hspace{0.3\linewidth}
    \left(\frac{M_\text{c} \sin i_\text{c}}{M_{\text{Jup}}}\right)\left(\frac{a_\text{c}}{10 \, \text{AU}}\right)^{-2} \sim 3.6 \tag{1}
,\end{equation}
where $a_\text{c}$ is the semimajor axis of the companion and $M_\text{c}$ is its mass. Using this estimation, HD 102888 c could, for instance, be a 14 M$_\text{Jup}$ brown dwarf on a 20 AU orbit, a 30 M$_\text{Jup}$ one on a 30 AU orbit or a 57 M$_\text{Jup}$ one on a 40 AU orbit. While this is only a rough estimate, this gives good hope that HD 102888 c could be a companion of HD 102888 of substellar nature. The full orbital parameters that we derived for this system can be found in Table \ref{tab:hd102888}.

\subsection{Update on the planetary system around HD 121056}
\begin{figure}[h]
    \centering
    \includegraphics[width=\columnwidth]{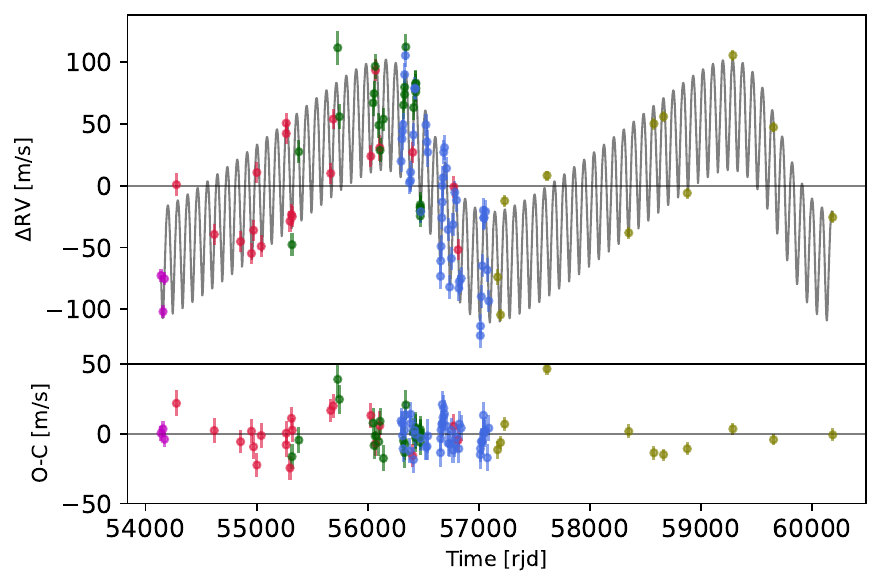}
    \caption{Same as Fig. \ref{fig:87816_timeseries} but for HD 121056. FEROS and CHIRON data are shown in dark green and dark blue, respectively.}
    \label{fig:121056_rvtseries}
\end{figure}
\begin{figure}[h]
    \centering
    \includegraphics[width=\columnwidth]{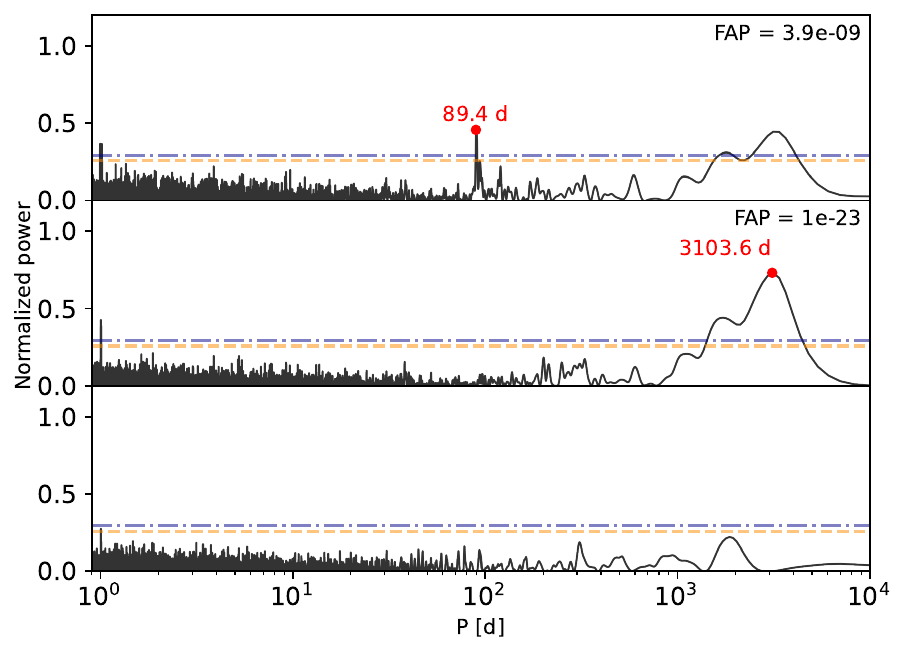}
    \caption{Same as Fig. \ref{fig:87816_periodo_planet} but for HD 121056.}
    \label{fig:121056_rv_periodo}
\end{figure}

\begin{table*}[h]

\centering
\caption{Comparison of the orbital parameters HD 121056 b and c from different studies.}
\begin{tabular}{lccccc}
\toprule
\toprule
Parameter & Unit&\citealp{jones_giant_2015} & \citealp{wittenmyer_pan-pacific_2015} & \citealp{feng_3d_2022} & This work \\
\midrule
$P_\text{b}$& [days] & 88.9 $\pm$ 0.1& 89.09 $\pm$ 0.11& $88.87^{+0.12}_{-0.12}$& $89.23^{+0.06}_{-0.06}$\\[3pt]
$K_\text{b}$& [m$\,$s$^{-1}$] & 45.5 $\pm$ 1.6& 47.9 $\pm$ 11.0& $46.66^{+1.16}_{-1.12}$& $44.68^{+1.73}_{-1.69}$\\[3pt]
$e_\text{b}$ && 0.05 $\pm$ 0.04&0.06 $\pm$ 0.04 & $0.06^{+0.03}_{-0.03}$& $0.13^{+0.04}_{-0.04}$\\[3pt]
$\omega_\text{b}$& [deg] & 138 $\pm$ 60& 300 $\pm$ 132& $101.94^{+35.6}_{-25.4}$         &$171.14^{+21.69}_{-21.99}$ \\[3pt]
$T_\text{P$_\text{b}$}$& [JD - 2 450 000] &2997.8 $\pm$ 16.7 &4810 $\pm$ 33 & $2989.08^{+78.7}_{-1.8}$    &$6471.79^{+5.33}_{-5.04}$ \\[3pt]
$m_\text{b}\sin i_\text{b}$& [M$_\text{Jup}$]&1.38 $\pm$ 0.15 &1.25 $\pm$ 0.04 & $1.57^{+0.07}_{-0.07}$& $1.15^{+0.04}_{-0.04}$\\
\midrule
$P_\text{c}$& [days] & 2131.8 $\pm$ 88.3& 2203 $\pm$ 486& $3923.66^{+616.25}_{-330.23}$& $3128.41^{+52.09}_{-47.13}$ \\[3pt]
$K_\text{c}$ &[m$\,$s$^{-1}$] & 69.0 $\pm$ 3.3& 81.9 $\pm$ 14.0&$62.29^{+7.6}_{-7.7}$ & $61.44^{+2.39}_{-2.34}$\\[3pt]
$e_\text{c}$& &0.17 $\pm$ 0.06 &0.18 $\pm$ 0.07 &$0.36^{+0.11}_{-0.09}$ & $0.30^{+0.03}_{-0.03}$ \\[3pt]
$\omega_\text{c}$ &[deg] & 166.5 $\pm$ 20.5& 205 $\pm$ 17& $321.35^{+11.6}_{-13.23}$         & $83.32^{+8.66}_{-8.22}$\\[3pt]
$T_\text{P$_\text{c}$}$& [JD - 2 450 000] &2684.1 $\pm$ 235.7 &3068 $\pm$ 405 & $1730.38^{+295.15}_{-560.66}$& $6574.13^{+50.57}_{-52.87}$\\[3pt]
$m_\text{c}\sin i_\text{c}$&[M$_\text{Jup}$] &5.98 $\pm$ 0.76 &6.14 $\pm$ 1.99 & 6.937$^{+2.05}_{-0.52}$& $4.97^{+0.22}_{-0.21}$\\
\bottomrule

\label{tab:121056comparison}

\end{tabular}
\tablefoot{For the solution of \citet{wittenmyer_pan-pacific_2015}, we only show here the results obtained with the "bootstrap solution" of the combined analysis (using FEROS, CHIRON, and UCLES) in the paper.}
\end{table*}

The presence of a planetary system around HD 121056 has already been reported by several teams in the past (\citealp{jones_planetary_2015}, \citealp{jones_giant_2015}, \citealp{wittenmyer_pan-pacific_2015} and \citealp{feng_3d_2022}). The detection of multi-planet systems around evolved stars remains rare, with fewer than 30 systems known to date\footnote{This number was obtained using the NASA exoplanet archive (\url{https://exoplanetarchive.ipac.caltech.edu/)} by restricting the detections to host stars with surface gravity $\log g$ $\leq$ 3.5 cm\,s$^{-2}$ and looking at systems with two or more planets.}. The system around HD 121056 is of particular interest, as it also features a short period planet, another rare discovery for giant stars\footnote{HD 121056 is one of two known planetary systems around a giant star with a period under 100 days and discovered by RV, alongside HD 33142 d (89.9 days, \citealp{trifonov_new_2022})}. While all the previous studies on HD 121056 agree on the period of the inner planet at approximately 89 days for a mass of about 1 M$_\text{Jup}$, different values for the period of the outer companion have been reported, as one can see in Table \ref{tab:121056comparison}.\par
\begin{figure}[h]
    \centering
    \includegraphics[width=\columnwidth]{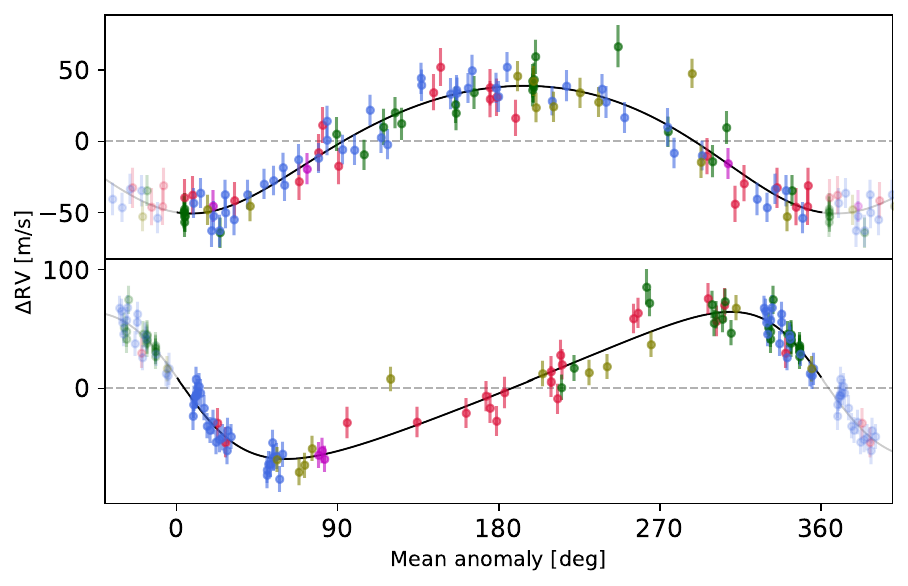}
    \caption{Top: Phase-folded RVs of HD 121056 b obtained with CORALIE, FEROS, and CHIRON. Bottom: Same but for HD 121056 c.}
    \label{fig:121056_phasefold}
\end{figure}

\begin{table}[h!]
\centering
\caption{Orbital parameters obtained for HD 121056 b and c.}
\begin{tabular}{lccc}
\toprule
\toprule
Parameter&Unit & HD 121056 b & HD 121056 c \\
\midrule
$P$&[days]  & $89.23^{+0.06}_{-0.06}$ & $3128.41^{+52.09}_{-47.13}$ \\[3pt]
$K$&[m$\,$s$^{-1}$]  & $44.68^{+1.73}_{-1.69}$ & $61.44^{+2.39}_{-2.34}$ \\[3pt]
$e$&  &  $0.13^{+0.04}_{-0.04}$ & $0.30^{+0.03}_{-0.03}$ \\[3pt]
$\omega$&[deg]  & $171.14^{+21.69}_{-21.99}$ & $83.32^{+8.66}_{-8.22}$ \\[3pt]
$\lambda_0$&[deg]  &  $74.17^{+2.78}_{-2.79}$ & $68.96^{+3.31}_{-3.12}$ \\[3pt]
$a$&[au]  & $0.42^{+0.0002}_{-0.0002}$ & $4.55^{+0.05}_{-0.05}$ \\[3pt]
$m_\text{p}$ sin $i$ & [M$_\text{Jup}$] & $1.15^{+0.04}_{-0.04}$ & $4.97^{+0.22}_{-0.21}$ \\[3pt]
\midrule
$\gamma_{\text{COR98}}$ &[m$\,$s$^{-1}$] & \multicolumn{2}{c}{$5723.65^{+7.53}_{-7.63}$}  \\[3pt]
$\gamma_{\text{COR07}}$ &[m$\,$s$^{-1}$] & \multicolumn{2}{c}{$5691.40^{+3.50}_{-3.63}$}  \\[3pt]
$\gamma_{\text{COR14}}$&[m$\,$s$^{-1}$] & \multicolumn{2}{c}{$5711.82^{+3.86}_{-3.97}$}  \\[3pt]
$\gamma_{\text{CHIRON}}$&[m$\,$s$^{-1}$] & \multicolumn{2}{c}{$12.82^{+3.26}_{-3.20}$}  \\[3pt]
$\gamma_{\text{FEROS}}$&[m$\,$s$^{-1}$] & \multicolumn{2}{c}{$-42.30^{+3.46}_{-3.54}$}  \\[3pt]
\midrule

$\sigma_{\text{CHIRON}}$ &[m$\,$s$^{-1}$]  & \multicolumn{2}{c}{ $2.39^{+2.54}_{-1.70}$}  \\[3pt]
$\sigma_{\text{FEROS}}$ &[m$\,$s$^{-1}$]  & \multicolumn{2}{c}{$3.41^{+3.22}_{-2.38}$}  \\[3pt]
$\sigma_{jit}$ &[m$\,$s$^{-1}$]  & \multicolumn{2}{c}{$10.63^{+1.13}_{-1.07}$}  \\[3pt]

\bottomrule
\end{tabular}
\label{tab:hd121056_params}
\tablefoot{The value of $\lambda_0$ is given for JD = 2456447.98}

\end{table}

Using the same approach as for the other targets studied in this work, we analyzed the RV time series of HD 121056 (Fig. \ref{fig:121056_rvtseries}) using a periodogram approach by fitting Keplerian models iteratively to the highest significant peak on the residuals (Fig. \ref{fig:121056_rv_periodo}). After subtracting our best two-planet model, the periodogram of the residuals does not show any significant peak. Our best-fit two-planet model can be seen in Fig. \ref{fig:121056_rvtseries}, where it is overplotted on the RV time series. The phase-folded RVs of HD 121056 b and c are shown in Fig. \ref{fig:121056_phasefold}. We looked for hints or stellar activity-related effects on the time series of FWHM, BIS, and H-alpha (Fig. \ref{fig:121056_activity}) but did not find any significant periodicities or correlation with the RVs.\par
We confirm the presence of the inner planet with a period of 89 days and mass $m_\text{b}\sin i_\text{b}$ = 1.2 M$_\text{Jup}$, and propose an update to the period of the outer companion. We find that HD 121056 c is a massive planet with minimum mass $m_\text{c}\sin i_\text{c}$ = 5 M$_\text{Jup}$ orbiting its host star in approximately 3130 days. Both planets are on relatively low-eccentricity orbits, with $e$ = 0.13 for HD 121056 b and $e$ = 0.3 for HD 121056 c. As our baseline of observation is much longer than that of previous studies, our data span a full orbit of HD 121056 c, allowing us to find a more accurate period of this second companion. All the orbital parameters of this system can be found in Table \ref{tab:hd121056_params}. A comparison with the results of other studies is given in Table \ref{tab:121056comparison}.

\section{Conclusion}
We have reported the detection of five new exoplanets orbiting evolved intermediate-mass stars.
\begin{itemize}
    \item HD 87816 b and HD 87816 c are two massive planets with minimum masses $m_\text{b}$$\sin i_b$ = 6.7 M$_\text{Jup}$ and $m_\text{c}$$\sin i_c$ = 12.2 M$_\text{Jup}$. The inner planet is on a very eccentric orbit, with an eccentricity $e$ = 0.78 and a period of 484 days, while the outer planet is on a less eccentric orbit with $e$ = 0.19 and a period of approximately 7600 days. 
    \item HD 94890 b and HD 94890 c are two massive planets from the same system, with minimum masses $m_\text{b}$$\sin i_b$ = 2.1 M$_\text{Jup}$ and $m_\text{c}$$\sin i_c$ = 8.9 M$_\text{Jup}$. They both have relatively long periods, 824 days for the inner planet and 2490 days for the outer. Both planets are on low-eccentricity orbits, $e_\text{b}$ = 0.22 and $e_\text{c}$ = 0.05.
    \item HD 102888 b is a 5.7 M$_\text{Jup}$ planet orbiting its giant host star in 252 days on a low-eccentricity orbit with $e$ = 0.11. After subtracting our best-fit model, we identified a remaining linear trend in the data, which is compatible with the presence of an unseen companion with a much longer period. Our analysis indicates that this companion is likely substellar and could be a more or less massive brown dwarf, depending on the distance at which it is orbiting its host star.
\end{itemize}\par
In addition to these three new planetary systems, we also confirmed the detection of HD 121056 b, which had already been reported by other teams, and provide an update on the orbital parameters of the second planet of the system HD 121056 c. We find an orbital period of 3128 days for this outer companion, which has a minimum mass $m_\text{c}\sin i_\text{c}$ = 5 M$_\text{Jup}$ on a moderately eccentric orbit with $e$ = 0.3. Regarding the inner companion, the orbital period that we find, approximately 89 days, is compatible with that of previous studies, though our estimate of the minimum mass is slightly lower, at $m_\text{b}\sin i_\text{b}$ = 1.2 M$_\text{Jup}$.\par
For all the reported new detections, we looked for hints of stellar activity by studying the time series of FWHM, BIS, and H-alpha. No correlation could be found between any of these quantities and the RV modulation that we observed, giving us confidence about the planetary-origin nature of our observations. After subtracting our best-fit models, the residuals of all systems exhibit weighted root mean squared values consistent with expectations for giant stars.
\label{sec:conclusion}
\section*{Data availability}
The RV data for all the targets discussed in this work are available in electronic form at the CDS via anonymous ftp to \url{cdsarc.u-strasbg.fr} (130.79.128.5) or via \url{http://cdsweb.u-strasbg.fr/cgi-bin/qcat?J/A+A/}. The data can also be downloaded and visualized on the DACE platform \url{https://doi.org/10.82180/dace-smvbg35u   }.
\begin{acknowledgements}
      We thank all the observers of the Swiss telescope at La Silla Observatory who contributed to the observations of this program over the past 19 years. This work was supported by the Swiss National Science Foundation (SNSF) under grant number 205010. This publication makes use of The Data \& Analysis Center for Exoplanets (DACE), which is a facility based at the University of Geneva (CH) dedicated to extrasolar planets data visualisation, exchange and analysis. DACE is a platform of the Swiss National Centre of Competence in Research (NCCR) PlanetS, federating the Swiss expertise in Exoplanet research. The DACE platform is available at \url{https://dace.unige.ch}. A.De. acknowledges the financial support of the National Centre of Competence in Research PlanetS supported by the Swiss National Science Foundation under grants 51NF40\_182901 and 51NF40\_205606. NCS acknowledges funding by the European Union (ERC, FIERCE, 101052347). Views and opinions expressed are however those of the author(s) only and do not necessarily reflect those of the European Union or the European Research Council. Neither the European Union nor the granting authority can be held responsible for them. This work was supported by FCT - Fundação para a Ciência e a Tecnologia through national funds by these grants: UIDB/04434/2020 DOI: 10.54499/UIDB/04434/2020, UIDP/04434/2020 DOI: 10.54499/UIDP/04434/2020. A.N. acknowledges support from the Swiss National Science Foundation (SNSF) under grant PZ00P2\_208945.
\end{acknowledgements}

\bibliographystyle{aa} 
\bibliography{cascades_1}

\begin{appendix} 
\section{Periodograms of FWHM, BIS, and H-alpha}
\begin{figure}[h]
\begin{subfigure}{.5\textwidth}
  \centering
  \includegraphics[width=.8\linewidth]{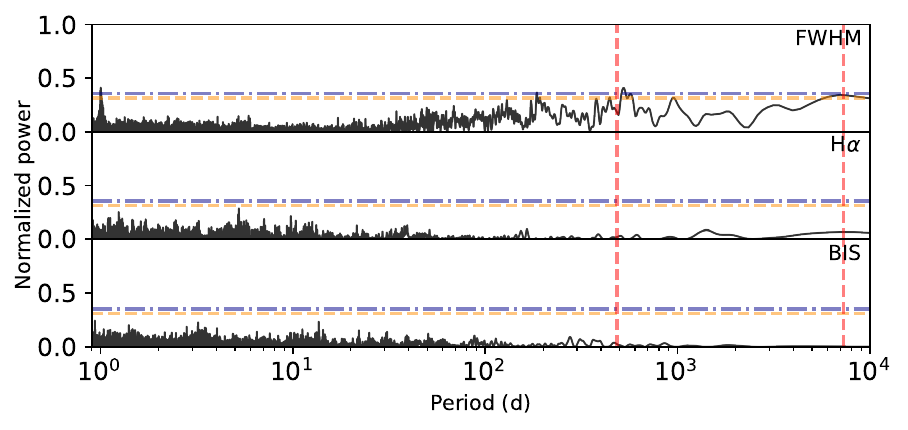}
  \caption{}
  \label{fig:87816_activity}
\end{subfigure}%
\begin{subfigure}{.5\textwidth}
  \centering
  \includegraphics[width=.8\linewidth]{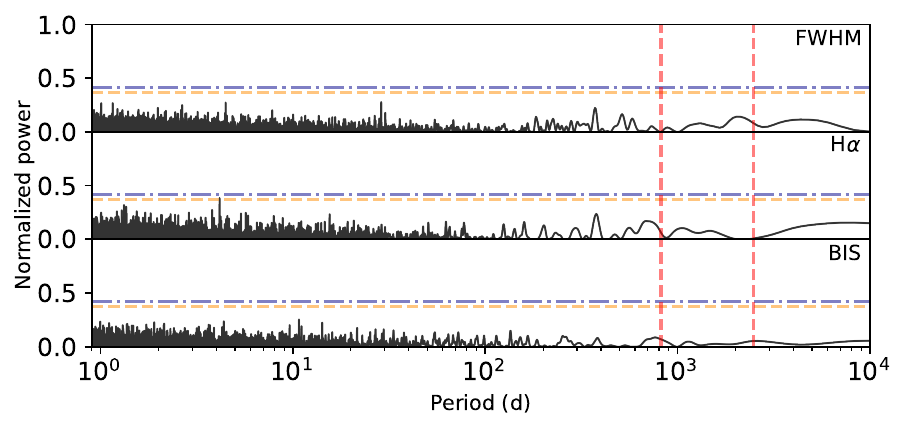}
  \caption{}
  \label{fig:94890periodo_activity}
\end{subfigure}

\begin{subfigure}{.5\textwidth}
  \centering
  \includegraphics[width=.8\linewidth]{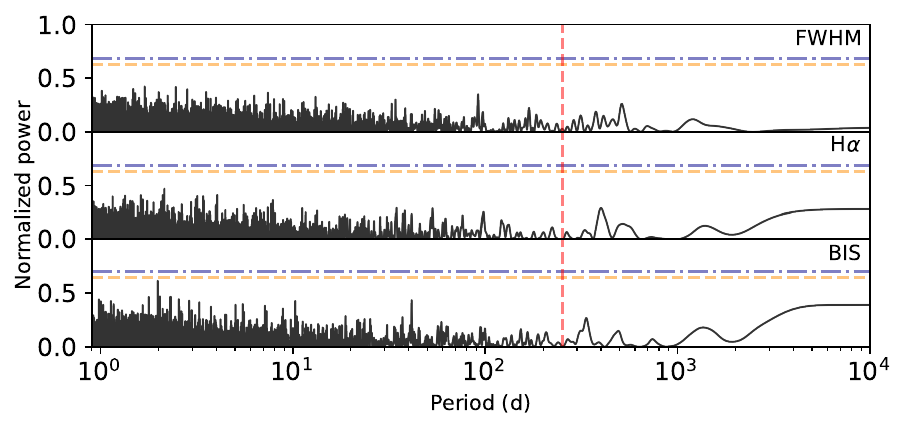}
  \caption{}
  \label{fig:102888_activityl}
\end{subfigure}%
\begin{subfigure}{.5\textwidth}
  \centering
  \includegraphics[width=.8\linewidth]{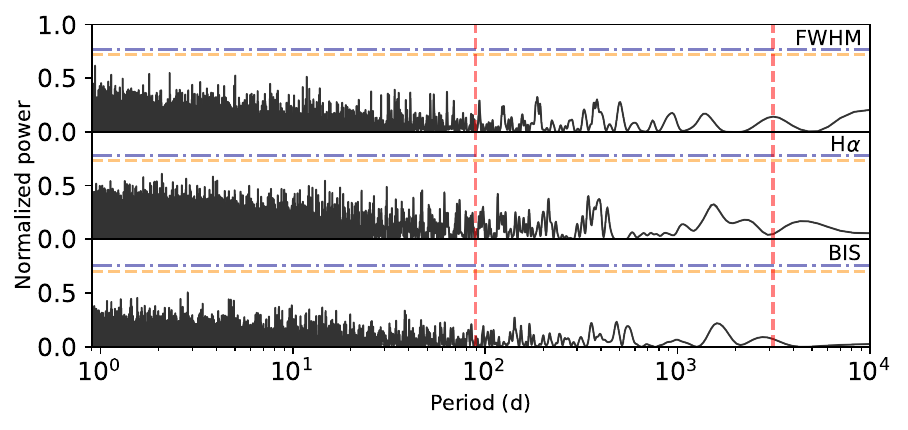}
  \caption{}
  \label{fig:121056_activity}
\end{subfigure}
\captionsetup{width = \textwidth}
\begin{minipage}{\textwidth}
\caption{Periodograms of the time series of the FWHM, BIS, and H-alpha index for the four targets of this work: HD 87816 (panel a), HD 94890 (panel b), HD 102888 (panel c), and HD 121056 (panel d). The horizontal lines show the different FAP levels: 1\% (dashed orange line) and 0.1\% (dashed-dotted blue line). The vertical red lines indicate the orbital periods of the detected exoplanets around each star.} 
\end{minipage}
\label{fig:periodos_activity}
\end{figure}

\clearpage

\section{Corner plot distribution of parameters}
\begin{figure}[H]
\sidecaption
    \centering
    \includegraphics[scale = 0.4]{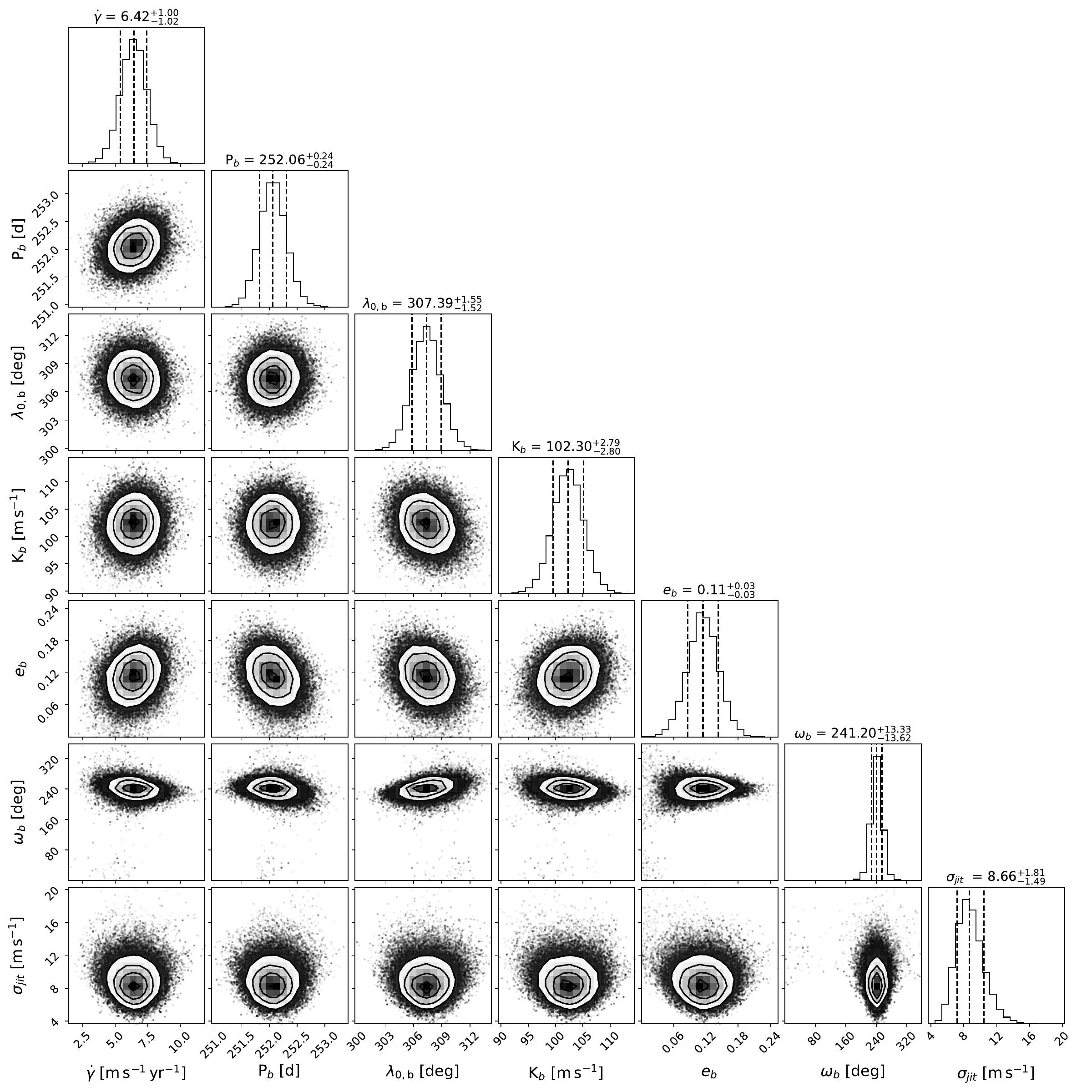}
    \caption{Posterior distributions of the orbital parameters of HD 102888 b obtained from the MCMC analysis.}
    \label{fig:corner_102888}
\end{figure}

\clearpage
\begin{figure*}[p]
    \centering
    \includegraphics[width=17cm]{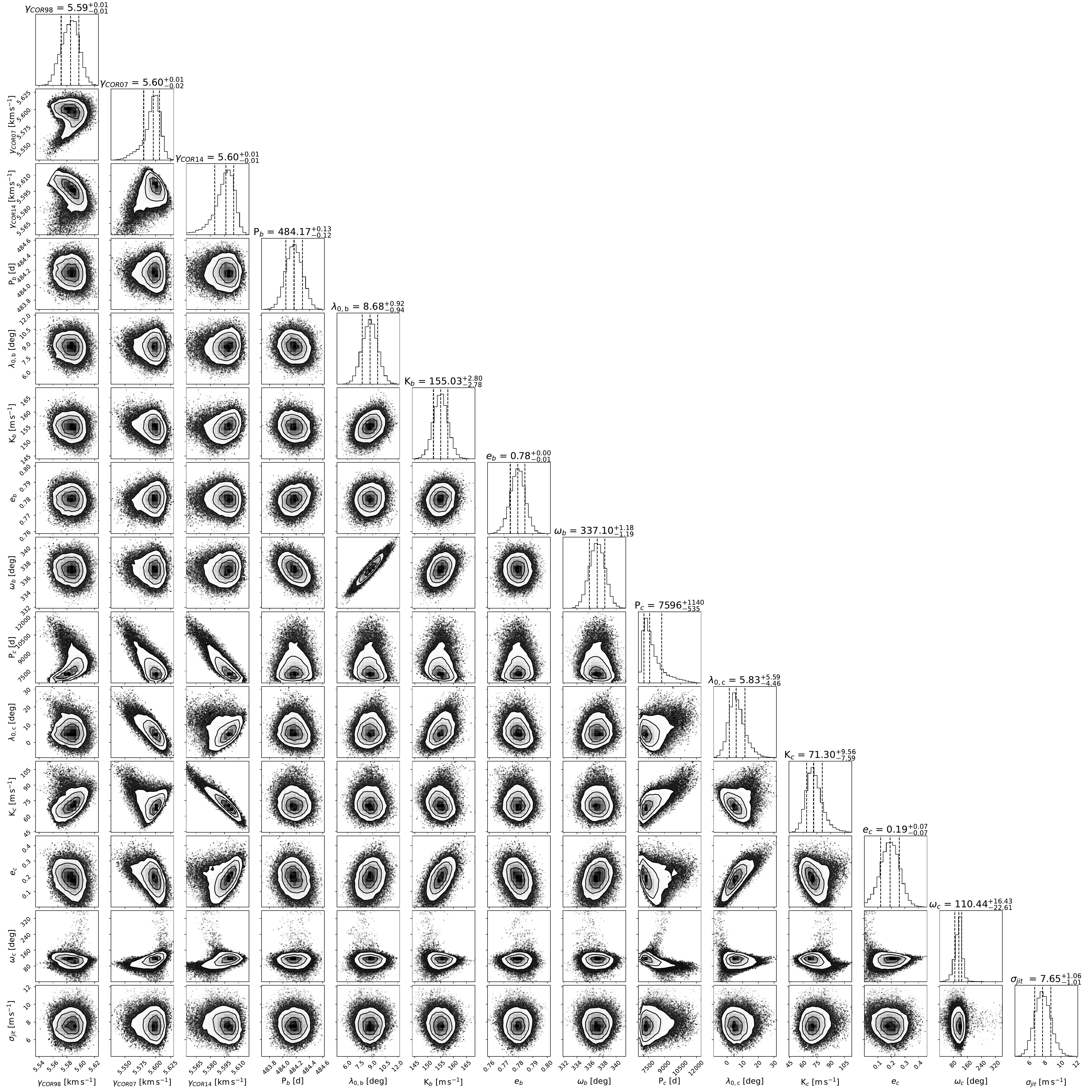}
    \caption{Same as Fig. \ref{fig:corner_102888} but for HD 87816 b and c.}
    \label{fig:corner_87816}
\end{figure*}
\clearpage
\begin{figure*}[p]
    \centering
    \includegraphics[width=\textwidth]{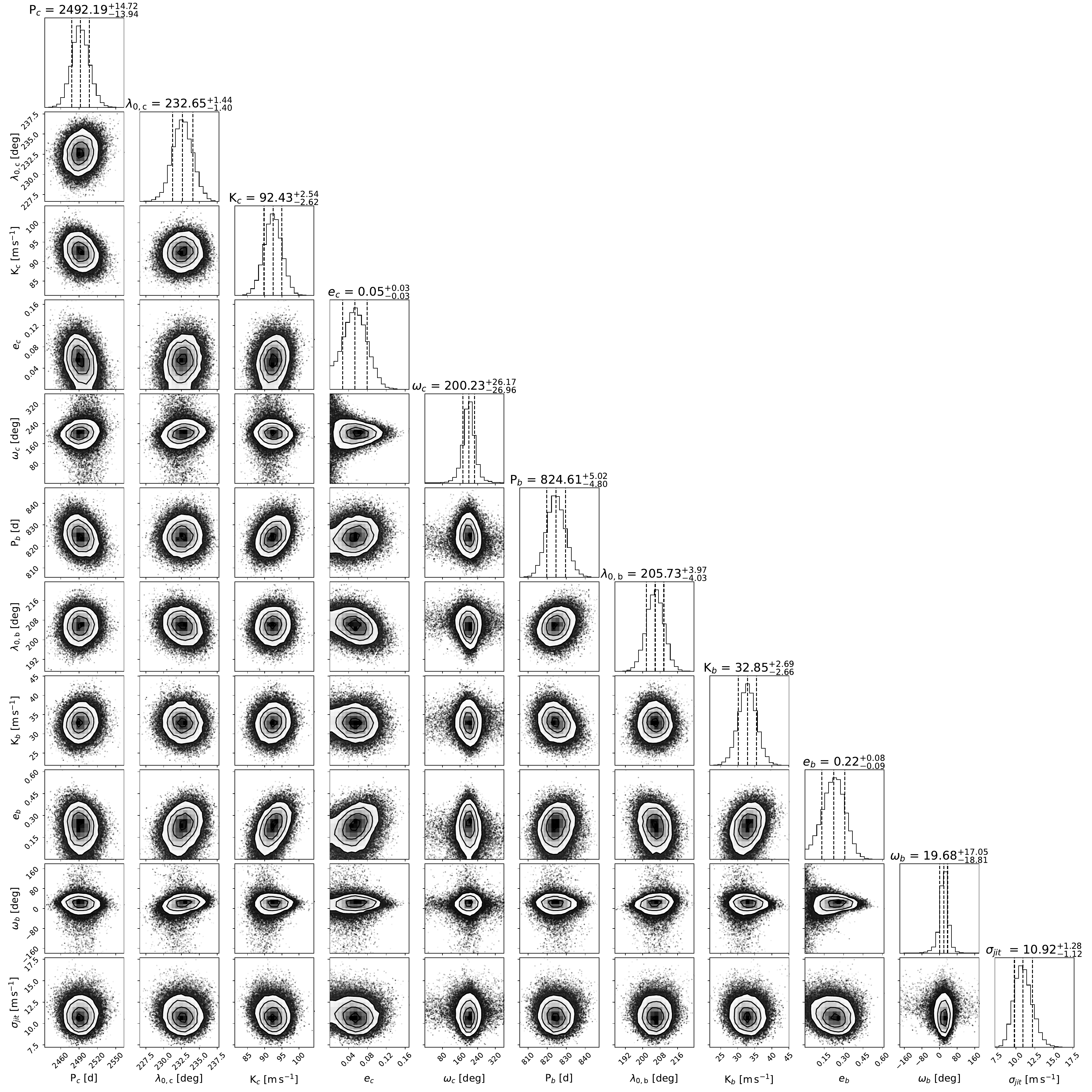}
    \caption{Same as Fig. \ref{fig:corner_102888} but for HD 94890 b and c.}
    \label{fig:corner_94890}
\end{figure*}
\clearpage

\begin{figure*}[p]
    \centering
    \includegraphics[width=\textwidth]{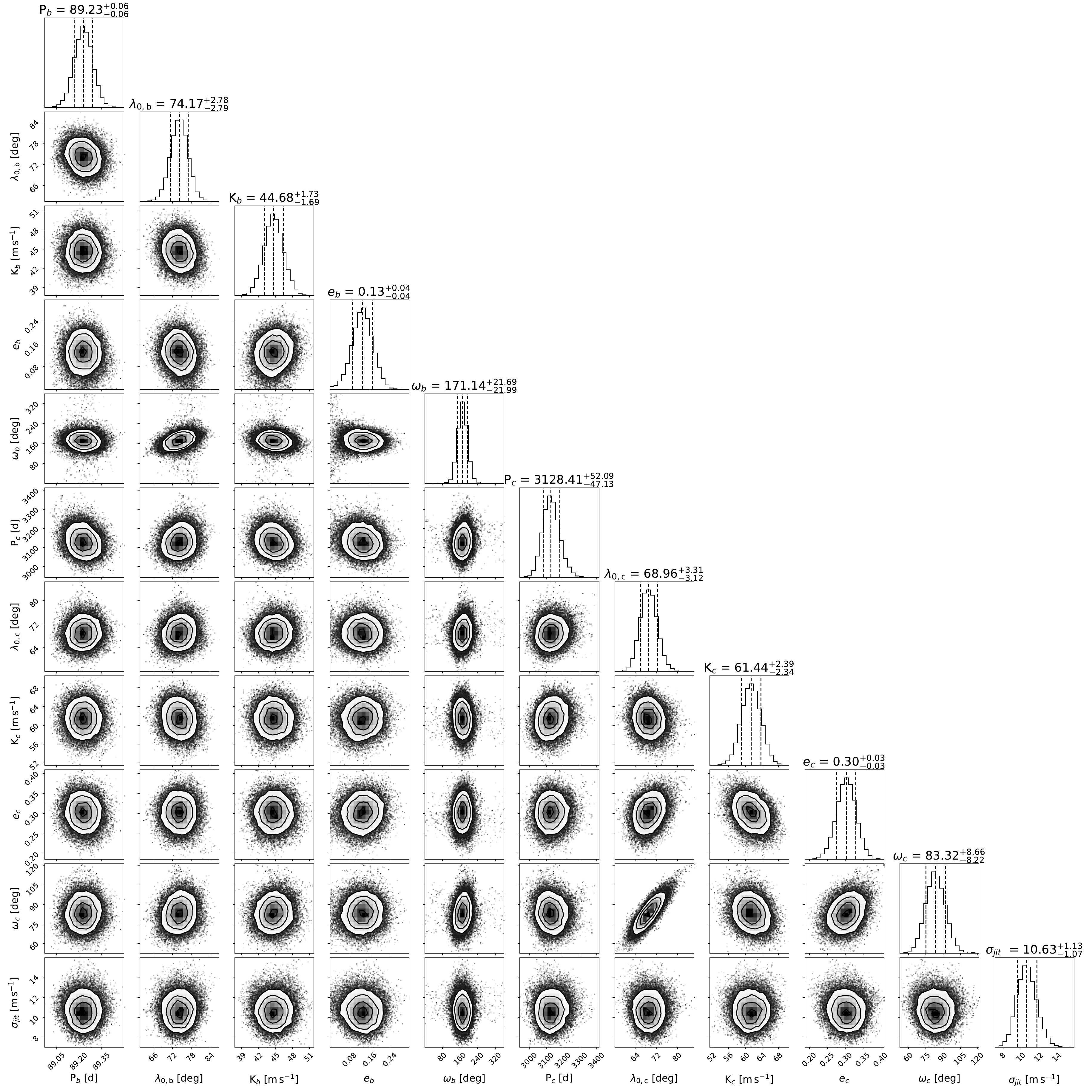}
    \caption{Same as Fig. \ref{fig:corner_102888} but for HD 121056 b and c.}
    \label{fig:corner_121056}
\end{figure*}

\end{appendix}
\end{document}